\documentclass[aip,amsmath,amssymb,preprint]{revtex4-1}

\usepackage{graphicx}
\usepackage{dcolumn}
\usepackage{bm}

\usepackage[utf8]{inputenc}
\usepackage[T1]{fontenc}
\usepackage{mathptmx}
\usepackage{etoolbox}
\usepackage{color}
\makeatletter
\def\@email#1#2{%
 \endgroup
 \patchcmd{\titleblock@produce}
  {\frontmatter@RRAPformat}
  {\frontmatter@RRAPformat{\produce@RRAP{*#1\href{mailto:#2}{#2}}}\frontmatter@RRAPformat}
  {}{}
}%
\makeatother
\begin{document}


\title[Photochemistry of Cyclobutanone]{The Photochemistry of Rydberg-Excited Cyclobutanone: Photoinduced Processes and Ground State Dynamics}
\author{J. Eng}
\affiliation{Chemistry, School of Natural and Environmental Sciences, Newcastle University, Newcastle upon Tyne, NE1 7RU, UK}
\email{julien.eng@newcastle.ac.uk}
\author{C. D. Rankine}
\affiliation{Department of Chemistry, University of York, York, YO10 5DD, UK}
\author{T. J. Penfold}
\email{tom.penfold@newcastle.ac.uk}
\affiliation{Chemistry, School of Natural and Environmental Sciences, Newcastle University, Newcastle upon Tyne, NE1 7RU, UK}

\date{\today}

\begin{abstract}
Owing to ring-strain, cyclic ketones exhibit complex excited-state dynamics with multiple competing photochemical channels active on the ultrafast timescale. While the excited-state dynamics of cyclobutanone after $\pi^{\ast}\leftarrow n$ excitation into the lowest-energy excited singlet state (S$_1$) has been extensively studied, the dynamics following 3$s\leftarrow n$ excitation into the higher-lying singlet Rydberg (S$_2$) state are less well understood. Herein, we couple quantum and excited-state trajectory surface-hopping molecular dynamics simulations to study the relaxation of cyclobutanone following 3s$\leftarrow n$ excitation and to predict the ultrafast electron diffraction scattering signal that we anticipate to arise from the relaxation dynamics that we observe. Our simulations indicate that relaxation from the initially-populated singlet Rydberg state occurs on the hundreds-of-femtosecond to picosecond timescale consistent with the symmetry-forbidden nature of the state-to-state transition involved. Once cyclobutanone has relaxed non-radiatively to the electronic ground state (S$_0$), the vibrationally hot molecules have sufficient energy to form multiple fragmentory products on the electronic ground-state surface including C$_2$H$_4$ + CH$_2$CO (\textbf{C2}; 20\%), and C$_3$H$_6$ + CO (\textbf{C3}; 2.5\%). We discuss the limitations of our simulations, how these may influence the outcome of the excited-state dynamics we observe, and -- ultimately -- the predictive power of the simulated experimental observable.
\end{abstract}

\maketitle

\section{Introduction}
\begin{figure}[ht]
    \centering
    \includegraphics[width=0.8\linewidth]{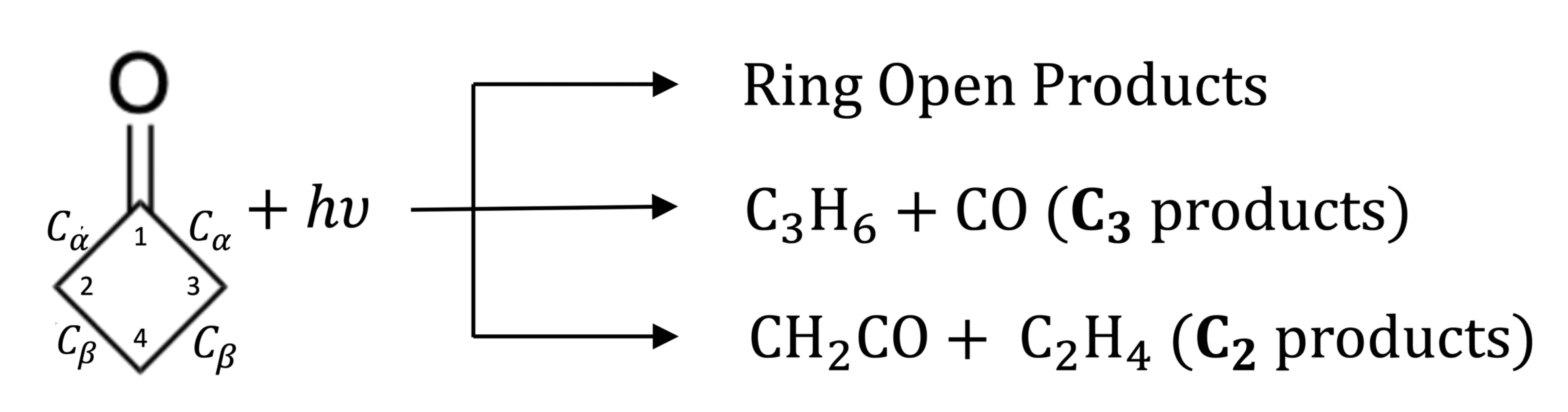}
    \caption{Schematic of cyclobutanone and the potential photoproducts identified in previous work.}
    \label{fig:photoproducts}
\end{figure}

Ketones, \textit{i.e.} organic compounds containing a carbonyl group, are among the simplest chromophores owing to their small size and low density of valence excited states. Consequently, their photochemistry has been intensively studied for several decades.\cite{norrish1936photodecomposition,turro1972molecular} Cyclic ketones form an important sub-class of these systems; despite their apparently equivalent simplicity, their photochemistry is often comparatively complex and features activity across multiple competing photochemical channels most commonly characterised by Norrish Type I processes. Here, upon excitation of an electron from the non-bonding orbital ($n$) on the carbonyl oxygen atom to the antibonding ($\pi^{\ast}$) molecular orbital of the carbonyl group, a carbon-carbon bond adjacent to the carbonyl (\textit{i.e.} the carbon-carbon bond to the $\alpha$ carbon, Figure \ref{fig:photoproducts}) cleaves, opening up the possibility of the formation of a variety of fragmentory products on the electronic ground-state surface after non-radiative relaxation. Amongst cyclic ketones, cyclobutanone has particular photochemistry owing to the high degree of ring strain in the cyclobutane ring (arising as a consequence of the small ring size).\cite{kao2020effects} The particular photochemistry of cyclobutanone has attracted significant interest from both an experimental \cite{denschlag1967mechanism,campbell1967energy,morton1973photochemical,hemminger1972fluorescence,lee1969tracer,turro1967molecular,kao2020effects,diau2001femtochemistry,kuhlman2012coherent,zhang1994jet,lee1971unusual,whitlock1971electronic,trentelman1990193,larsen2017coherent} and theoretical \cite{xia2015excited,liu2016new,hopkinson1988molecular,momicchioli1975structure,dalton1970photoreactivity,chen2004photochemical,kuhlman2012symmetry,kuhlman2013quantum} perspective. 

Many of these previous studies have aimed at elucidating the photochemistry and, in particular, the excited-state dynamics arising from excitation at the weak S$_1\leftarrow$S$_0$ absorption band (\textit{ca.} 330-240nm) associated with the symmetry-forbidden $\pi^{\ast}\leftarrow n$ transition. Early ultrafast spectroscopy carried out by Diau \textit{et al.}\cite{diau2001femtochemistry} reported that, upon excitation at 307 nm (\textit{i.e.} slightly above energy of the $\pi^{\ast}\leftarrow n$ transition by \textit{ca.} 2 kcal.mol$^{-1}$) $\alpha$-cleavage occurs over a timescale of \textit{ca.} 5 ps, driven initially by the C=O out-of-plane wagging, cyclobutane ring puckering, and C=O stretching modes. Following $\alpha$-cleavage, a channel along which an S$_{1}$/S$_{0}$ conical intersection (CI) is encountered and \textit{via} which non-radiative relaxation to the electronic ground state can proceed, cyclobutanone can undergo substantial structural changes that lead to the production of either i) a vibrationally-hot S$_{0}$ species, ii) a diradical intermediate that fragments to yield C$_2$H$_4$ + CH$_2$CO (\textbf{C2} products), or iii) the formation of C$_3$H$_6$ + CO (\textbf{C3} products). The proposed fragmentation mechanism is consistent with the observed wavelength dependence of the \textbf{C3}:\textbf{C2} product ratios;\cite{denschlag1967mechanism,lee1969tracer,lee1971unusual,hemminger1972fluorescence,breuer1971fluorescence} this ratio is reported to be 0.5 on excitation at 313 nm, and increases to 0.8 at 248 nm and 1.0 at 200 nm.\cite{lee1971unusual,tang1976laser} At wavelengths longer than 315 nm, there is a marked increase in the \textbf{C3}:\textbf{C2} product ratio; it is reported to be 2.0 at 326 nm and as high as 7.0 at 344 nm.\cite{hemminger1972fluorescence,tang1976laser} The increase is indicative of an alternative mechanism in the S$_1$ state which becomes operational at longer wavelengths and it is proposed to involve photochemistry on the triplet manifold \cite{tang1976laser} as there is insufficient energy available to overcome the barrier encountered along the $\alpha$-cleavage channel and reach the S$_{1}$/S$_{0}$ CI.  This mechanistic picture is largely maintained in solution as demonstrate by Kao \textit{et al.}\cite{kao2020effects} who observed Norrish Type I $\alpha$-cleavage on the sub-picosecond timescale, a consequence of excess vibrational energy available from UV-photon absorption from the electronic ground state. Kao \textit{et al.}\cite{kao2020effects} proposed that the observed wavelength dependence arose from the need to overcome a barrier on the S$_1$ surface (\textit{e.g.} the barrier encountered along the $\alpha$-cleavage channel). In addition, while triplet states have be invoked in the Norrish mechanism,\cite{lee1971unusual} Kao \textit{et al.}\cite{kao2020effects} reported that, even with sufficient excess vibrational energy available, triplet states may still play a mechanistic role although only in indirect channels active on timescales longer than 500 ps.

The second absorption band of cyclobutanone occurs between 206 to 182 nm and has been assigned to the $3s \leftarrow n$ transition into the second electronically-excited singlet state (S$_2$); a state that exhibits Rydberg character.\cite{whitlock1971electronic} Trentelman \textit{et al.}\cite{trentelman1990193} have studied the 193 nm photolysis of cyclobutanone and have reported that 57\% of the electronically-excited-state species form the \textbf{C3} products, 30\% form the \textbf{C2} products in the electronically excited state, and 13\% form the \textbf{C2} products in the (hot) electronic ground state. Importantly, the observations of Trentelman \textit{et al.}\cite{trentelman1990193} suggest that the formation of the \textbf{C3} products is potentially a slow process as it requires intersystem crossing (ISC) onto the triplet manifold. However, the proposed photochemical mechanism assumes an S$_{1}$ intermediate that survives long enough to undergo thermalisation and, therefore, produce a statistical partitioning of products; this assumption does not necessarily hold given the speed at which S$_{0}$ $\leftarrow$ S$_{1}$ relaxation is expected to take place. Kuhlman \textit{et al.}\cite{kuhlman2012coherent,kuhlman2013pulling} have studied the 200 nm photolysis of cyclobutanone, comparatively, and reported conversion from the S$_2$ to the S$_1$ associated with two time constants of \textit{ca.} 350 and \textit{ca.} 750 fs. Using time-resolved mass spectrometry (TR-MS), the authors were able to resolve two fragments: the parent, and a fragment with a mass-to-charge ratio $m/z = 42$, corresponding to H$_{2}$C--C=O, both exhibiting similar electronically-excited-state decay constants. The authors also reported a blue-shift in the photoelectron spectrum, arising from a coherent oscillatory motion assigned to a low frequency ring puckering mode with a frequency of 35 cm$^{-1}$, promoted by removal of an electron from an oxygen $n$ orbital which leads to a relaxation of the sp$^2$ hybridization of the carbonyl carbon due to mixing with components of the bonding orbitals in the carbonyl group as well as a relaxation of the adjacent C-C bonds.

From the perspective of theory, Xia \textit{et al.}\cite{xia2015excited} performed a quantum chemistry investigation optimized minima, transition states, MECIs, and relaxed two-dimensional S$_1$ and T$_1$ potential energy surfaces using high-level quantum chemical calculations. On the basis of these calculations, Xia \textit{et al.} proposed that ring opening predominately occurs in the S$_{1}$ state through an accessible S$_{1}$/S$_{0}$ MECI. Liu \textit{et al.}\cite{liu2016new} extended this work by performing dynamics simulations within the \textit{ab initio} multiple spawning (AIMS)\cite{ben2000ab,curchod2018ab} framework following S$_{1}$ photoexcitation. They observed that relaxation occurred primarily through the S$_{1}$/S$_{0}$ MECI associated with fission of the first $\alpha$-C-C bond, which fell into the region with the $\beta$-C-C bond is $\sim$1.6 \mbox{\AA}, and the $\alpha$-C-C bond between 2.5-3.5 \mbox{\AA}. In comparison they found that passage through intersections associated with the formation of the \textbf{C2} products, occurred when the $\beta$-C-C bond was around 4 \mbox{\AA}, but this constitutes only 15\%. They reported an excited state lifetime in the S$_{1}$ state of \textit{ca.} 500 fs and, on this basis, proposed that non-ergodic behavior was the driver of the change in \textbf{C3}:\textbf{C2} branching ratio as a function of excitation wavelength. In contrast, Kuhlman \textit{et al.}\cite{kuhlman2013quantum} developed a four-state, five-dimensional model Hamiltonian and carried out quantum dynamics to study the S$_{1}$ $\leftarrow$ S$_{2}$ relaxation dynamics. They concluded that S$_{1}$ $\leftarrow$ S$_{2}$ relaxation involves specific nuclear modes included in the model Hamiltonian (including the cyclobutane ring-puckering, carbonyl out-of-plane deformation, symmetric and asymmetric C-CO-C stretches, and carbonyl stretching modes) that couple the S$_{2}$ and S$_{1}$ states and promote population transfer on the picosecond timescale. While informative, these potentials are built within a harmonic normal mode representation and are unable to address satisfactorily large amplitude nuclear motions associated with the formation of photoproducts. 

Despite the concerted efforts of experiment and theory, there remain a significant number of open questions regarding the photochemistry and photochemical dynamics of cyclobutanone which evolve post-photoexcitation to the S$_{2}$ Rydberg ($3s \leftarrow n$) state. Nonetheless, the emergence of modern light sources \cite{filippetto2022ultrafast,pellegrini2016physics} is facilitating the study of ultrafast structural dynamics using both X-rays and electrons\cite{chergui2009electron,yang2020simultaneous,katayama2023atomic,minitti2015imaging} with ever-increasing spatial and temporal resolution and providing the potential for new and increasingly detailed insights into complex photochemical processes such as these. These developments are bringing into focus a crucial question (which is the focus of the present Special Issue to which this Article contributes): \textit{how accurate are modern nonadiabatic excited-state molecular dynamics simulations really?}

In this Article, we combine quantum- and excited-state trajectory surface-hopping molecular dynamics simulations at the LR-TDDFT(PBE0) and ADC(2) levels of theory to explore the relaxation dynamics of cyclobutanone post-photoexcitation to the S$_{2}$ Rydberg ($3s \leftarrow n$) state. Our excited-state molecular dynamics simulations are subsequently used predict the scattering signals for an ultrafast electron diffraction experiment in reciprocal and real space to establish how the complex photochemical dynamics of cyclobutanone might manifest as an experimental observable. In such an ultrafast electron diffraction experiment, free cyclobutanone molecules introduced \textit{in vacuo} will be photoexcited at 200 nm [\textit{i.e.} into the S$_{2}$ Rydberg ($3s \leftarrow n$) state] and probed \textit{via} electron scattering at a sequence of temporal delays to image directly the evolving excited-state structural dynamics.

\section{Theory and Computational Details}

\subsection{Quantum Chemistry}

All quantum chemical calculations were carried out using Turbomole (v7.4).\cite{turbomole,turbomole2} The quantum chemical calculations on the electronic ground and excited states used density functional theory (DFT) and linear-response time-dependent DFT (LR-TDDFT), respectively, with the PBE0 functional\cite{pbe0} and the aug-cc-pVDZ basis set.\cite{BS1,BS2} The Tamm-Dancoff\cite{TDA} approximation (TDA) was used throughout. Vibrational analysis were performed to confirm the absence of imaginary frequencies at the ground state minimum. Minimum-energy conical intersections (MECIs) were optimised at the LR-TDDFT(PBE0) level using Turbomole (v7.4) coupled with an external (penalty-function-based) optimiser.\cite{rankine_optci}. Linear reaction channels (produced \textit{via} linear interpolation in internal coordinates; LIIC) between critical geometries and these CIs are available in the SI alongside benchmarks of the electronic structure and basis set. 

Additional quantum chemical calculations were carried out at critical geometries using the second-order algebraic diagrammatic construction scheme [ADC(2)] and \textit{n}-electron valence state perturbation theory (NEVPT2). All ADC(2) calculations were carried out using Turbomole (v7.4);\cite{turbomole,turbomole2} all NEVPT2 calculations were carried out using ORCA.\cite{orca_1,orca_2} 

\subsection{Vibronic Coupling Hamiltonian and Quantum Dynamics}

The model Hamiltonian developed is based upon the vibronic coupling approximation\cite{Koppel2004,Koppel2009} and is expressed as:

\begin{equation}
    \mathbf{H}=\left(T_N+V_0\right)\mathbf{1} + \mathbf{W}^{dia.} + \mathbf{W}^{SOC} 
\end{equation}

The electronic diabatic Hamiltonian elements, $W_{n,n}$, are obtained by expanding $\mathbf{W}-V_0\mathbf{1}$ the diabatic potential as a Taylor series around a reference nuclear geometry, $Q_0$, here taken as the Franck-Condon (\textit{e.g.} S$_{0}$ minimum-energy) geometry. $V_0$ is a reference potential, here defined as a set of harmonic potentials with vibrational frequencies $\omega_i$ along dimensionless normal coordinates $Q_i$. In this case, the Hamiltonian elements are expressed as:

\begin{equation}
W_{n,m}-V_o\delta_{nm}=\epsilon_n\delta_{nm} + \sum_i^{3N-6}\left. \frac{\partial W_{n,m}}{\partial Q_i}\right|_{Q_0}Q_i + \frac{1}{2!}\sum_{i,j}^{3N-6}\left. \frac{\partial^2 W_{n,m}}{\partial Q_i \partial Q_j}\right|_{Q_0}Q_iQ_j + \frac{1}{3!}\left.\frac{\partial^3 W_{n,m}}{\partial Q_i \partial Q_j \partial Q_k}\right|_{Q_0}Q_iQ_jQ_k+\cdots
\end{equation}

where $\delta_{nm}$ is the Kronecker delta. $Q_i$ denotes the $3N-6$ dimensionless normal coordinates related to the normal modes of vibration where $N$ is the number of atoms. $V_0$, under the harmonic approximation, is expressed as:

\begin{equation}
V_0=\frac{1}{2}\omega_iQ_i^2
\end{equation} 

and the kinetic energy operator, $\hat{T}_{N}$, takes the form:

\begin{equation}
\hat{T_N}=-\sum_i\frac{1}{2}\omega_i \frac{\partial^2}{\partial Q_i^2}
\end{equation}

Due to the anharmonicity of the potential energy surface, the Taylor series expansion is carried out up to fourth order.\cite{penfold2009model,penfold2012quantum} Obtaining the expansion coefficients for the Hamiltonian is simplified for the present system \textit{via} use of symmetry; the minimum-energy S$_{0}$ geometry of cyclobutanone is C$_{s}$ symmetric and, consequently, first- and second-order couplings are only allowed if the following selection rules are satisfied:

\begin{equation}
    \Gamma_n\otimes\Gamma_{Q_a}\otimes\Gamma_n\supset A'
\end{equation}
\begin{equation}
    \Gamma_n\otimes\Gamma_{Q_a}\Gamma_{Q_b}\otimes\Gamma_n\supset A' ,
\end{equation}

All expansion coefficients were obtained using a development version of the in-house-developed VCMaker\cite{GAP,tango} software, available at Ref. \cite{VCMaker}.

Quantum dynamics simulations over the multi-dimensional potential energy surface(s) were carried out using the multiconfigurational time-dependent Hartree (MCTDH)\cite{beck2000multiconfiguration,meyer2009multidimensional} approach as implemented in Quantics.\cite{quantics} The initial wavefunction for the electronic ground state was built using one-dimensional harmonic oscillator functions with zero initial momentum and vertically excited into the S$_{2}$ state at the Franck-Condon geometry ($Q$ = 0). The complete computational details are provided in the SI and ensured convergence for the 2 ps of dynamics presented. 

\subsection{Excited-State Trajectory Surface-Hopping Molecular Dynamics}

\textit{Ab initio} excited-state trajectory surface-hopping molecular dynamics were performed using Newton-X (v2.4).\cite{Newtonx1,Newtonx2} The potential energy surfaces, derivatives, and respective couplings were computed on-the-fly using Turbomole (v7.4)\cite{turbomole,turbomole2} at two separate levels of theory [LR-TDDFT(PBE0) and ADC(2)] in two separate sets of simulations. The trajectories were propagated using the velocity Verlet algorithm\cite{Verlet1,Verlet2} for a maximum of 5 ps ($t_{\mathrm{max}} = 5000$ fs) with a time step of 0.5 fs ($dt = 0.5$ fs). The state-to-state transitions were simulated using the Hammes-Schiffer-and-Tully fewest-switches algorithm\cite{TSH} with the state-to-state coupling estimated \textit{via} the time-dependent Baeck-An\cite{baeckan} coupling scheme. The excited-state trajectory surface-hopping molecular dynamics simulations were initiated by simulated vertical projection of a set of initial conditions, generated according to a Wigner distribution with a temperature of 100K, from the electronic ground state (S$_{0}$) into the S$_2$ Rydberg ($3s \leftarrow n$) state.

To avoid instabilities at/around the S$_1$/S$_0$ crossing seam, S$_{0}$ $\leftarrow$ S$_{1}$ surface hops were forced when the S$_{1}$/S$_{0}$ energy gap fulfilled the criterion $\Delta{E}_{\mathrm{{S_1-S_0}}} < 0.1$ eV.

\subsection{Electron Diffraction Simulations}

Throughout this work, we present the ultrafast electron diffraction scattering signal as a modified electron scattering intensity ($sM(\mathbf{s})$) as computed under the independent atom model (IAM). This presentation is used to enhance the oscillations in the scattering signal associated with the molecular interference terms and suppress the rapid drop-off in scattering signal intensity as a function of $s$ from the elastic scattering amplitude. The modified electron scattering intensity is given by: 

\begin{equation}
sM(\mathbf{s}) = \dfrac{I_{\mathrm{mol.}}(\mathbf{s})}{I_{\mathrm{at.}}(\mathbf{s})}\mathbf{s}
\end{equation}

where $\mathbf{s}$ is the momentum transfer, or scattering, vector. $I_{\mathrm{at.}}(\mathbf{s})$ is the atomic scattering term, given by: 

\begin{equation}
I_{\mathrm{at.}}(\mathbf{s}) = \sum_i^{N} \vert f_i(\mathbf{s}) \vert^2
\end{equation}

and $I_{\mathrm{mol.}}(\mathbf{s})$ is the molecular scattering term, expressed as a sum of interference terms, and given by:

\begin{equation}
I_{\mathrm{mol.}}(\mathbf{s}) = \sum_i^{N} \sum_{j \neq i}^{N} \vert f_i(\mathbf{s}) \vert \vert f_j(\mathbf{s}) \vert \dfrac{sin(\mathbf{s}r_{ij})}{\mathbf{s}r_{ij}}
\end{equation}

where $f_i(\mathbf{s})$ and $f_j(\mathbf{s})$ are the elastic scattering amplitudes for atoms $i$ and $j$, respectively, and $r_{ij}$ is the internuclear distance between atoms $i$ and $j$.

$sM(\mathbf{s})$ can be transformed into a pair-distribution function (PDF; \textit{i.e.} from reciprocal space into real space) using a sine transform:

\begin{equation}
\mathrm{PDF}(\mathbf{r}) = \int_{0}^{s_{\mathrm{max}}} sM(\mathbf{s}) \sin(\mathbf{sr})e^{-k\mathbf{s}^2} ds
\end{equation}

where $s_{max}$ is the maximum momentum transfer in the data, $r$ is the internuclear distance between atom pairs, and $k$ is a damping factor used to drive down smoothly the contribution(s) at high $s$ to zero. A damping factor of $k = 0.03$ was used throughout.

The focus of this work is upon the ultrafast excited-state dynamics of cyclobutanone and, accordingly, the ultrafast electron diffraction scattering signal is presented as a transient (\textit{i.e.} as an excited state $-$ ground state '\textit{difference}') signal where appropriate under the nuclear ensemble model. The ground-state signal used to generate the transient is predicted from ground-state molecular dynamics, \textit{i.e.} a nuclear ensemble of equilibrated configurations pre-photoexcitation. No scaling for, \textit{e.g.}, excitation percentage/photolysis yield has been carried out. 

\section{Results}

\subsection{Characterizing Critical Points on the Ground- and Electronically-Excited Potential Energy Surfaces}

Table \ref{tab:es} shows the energies of the low-lying singlet (S$_{n}$; $n = [1..5]$) and triplet (T$_{n}$; $n = [1..5]$) electronically-excited states of cyclobutanone. Figure S1 shows the molecular orbitals involved in the electronic transitions. The symmetry point group of cyclobutanone (C$_s$ at the S$_{0}$ minimum-energy geometry) is used to characterise the symmetries of the electronically-excited states (an important facet of the generation and interpretation of the model Hamiltonian used in the following section). 

At the S$_{0}$ minimum-energy geometry, the S$_1$ and S$_2$ excited states are located 4.21 and 5.99 eV, respectively, above the electronic ground state at the ADC(2)/aug-cc-pVTZ level of theory, and at 4.33 and 6.10 eV, respectively, above the electronic ground state at the LR-TDDFT(PBE0)/aug-cc-pVTZ level of theory. Both levels of theory give good agreement with the experimental absorption spectrum recorded by Diau \textit{et al.},\cite{diau2001femtochemistry} in which absorption bands are observed at \textit{ca.} 4.2 and \textit{ca.} 6.2 eV. At all levels of theory the S$_{1}$ and S$_{2}$ states have $\pi^{\ast}\leftarrow n$ (LUMO $\leftarrow$ HOMO) and $3s\leftarrow n$ (LUMO+1 $\leftarrow$ HOMO) character, respectively. The Rydberg character of the S$_{2}$ state requires a larger basis set to describe accurately its energy and electronic structure, while -- in contrast -- the valence S$_{1}$ state exhibits little to no dependence on the basis set (Table S1). Both the S$_{2}$ and S$_{1}$ states are of A" symmetry; direct first-order vibronic coupling between these two states is consequently forbidden, and this can be expected to slow down the rate of internal conversion between the two states. 

The ADC(2) calculations clearly give very good agreement with experimental observations at the Franck-Condon geometry. However, previous work has highlighted a limitation of this approach for studying non-radiative pathways of carbonyl-containing molecules \cite{marsili2021caveat} as it predicts an artificial S$_1$/S$_0$ crossing along C=O vibrational arising from a n$\rightarrow\pi^{\ast}$ which is too shallow combined with a ground state which destabilises too rapidly. To address this, we calculate the potential along the C=O stretching normal mode, which is shown in Figure S2 calculated using LR-TDDFT(PBE0), ADC(2) and NEVPT2(12,12) levels of theory. This shows good agreement between the LR-TDDFT(PBE0) and NEVPT2(12,12) simulations, and although the ground state and S$_1$ surfaces become closer, they do not cross. In contrast, as expected from ref. \cite{marsili2021caveat}, there is a clear crossing between the two surfaces in ADC(2), which corresponds to the C=O bond length of $\sim$1.6 \mbox{\AA}. This low lying accessible crossing point is likely to distort the excited state dynamics crossing from the S$_1$ to ground state surface and indeed such dynamics are presented in the supplementary material. 

\begin{table}[ht]
    \centering
    \begin{tabular}{|c||c|c||c|c||}
         \hline
         State & \multicolumn{2}{|c||}{ADC(2)} & \multicolumn{2}{|c||}{LR-TDDFT(PBE0)}  \\
   \hline
    & Energy / eV & Character (Irrep.) & Energy / eV & Character (Irrep.)  \\
   \hline
        GS      &  0.00 &  ---   (A')  & 0.00 & ---   (A')  \\   
        S$_1$   &  4.21 &  H$\rightarrow$L   (A")  & 4.33 & H$\rightarrow$L   (A")  \\
        S$_2$   &  5.99 &  H$\rightarrow$L+1 (A")  & 6.10 & H$\rightarrow$L+1 (A") \\
        S$_3$   &  6.53 &  H$\rightarrow$L+3 (A")  & 6.67 & H$\rightarrow$L+3 (A") \\
        S$_4$   &  6.64 &  H$\rightarrow$L+2 (A')  & 6.77 & H$\rightarrow$L+2 (A') \\
        S$_5$   &  6.68 &  H$\rightarrow$L+4 (A")  & 6.78 & H$\rightarrow$L+4 (A") \\
        T$_1$   &  3.87 &  H$\rightarrow$L   (A")  & 3.73 & H$\rightarrow$L   (A")\\
        T$_2$   &  5.95 &  H-1$\rightarrow$L (A')  & 5.98 & H-1$\rightarrow$L (A') \\
                &       &  H-4$\rightarrow$L       &        & H-4$\rightarrow$L \\
        T$_3$   &  6.32 &  H$\rightarrow$L+1 (A")  & 6.02 &H$\rightarrow$L+1   (A")  \\
        T$_4$   &  6.50 &  H$\rightarrow$L+3 (A")  & 6.61 & H$\rightarrow$L+3 (A")\\
        T$_5$   &  6.62 &  H$\rightarrow$L+2 (A')  & 6.63 & H$\rightarrow$L+2 (A') \\
 \hline
    \end{tabular}
    \caption{Summary of electronic energies, characters, and symmetries of the ground- (GS) and electronically-excited (S$_{n}$/T$_{n}$; $n = [1..5]$) states of cyclobutanone evaluated at the S$_{0}$ minimum-energy geometry and at the ADC(2)/aug-cc-pVTZ and LR-TDDFT(PBE0)/aug-cc-pVDZ levels of theory. Comparative tables for evaluations at the S$_1$- (Table S3) and S$_2$-state (Table S4) minimum-energy geometries are presented in the SI. The relevant molecular orbitals are shown in the SI. The HOMO is designated as H; the LUMO is designated as L.}
    \label{tab:es}
\end{table}

The S$_1$ state minimum-energy geometry is reached from the Franck-Condon point \textit{via} an out-of-plane puckering and slight elongation of the C=O bond (1.21 to 1.26 \mbox{\AA}); the C-C bonds in the cyclobutane ring remain almost unchanged. These structural changes destabilise the electronic ground state (Table S3); it is increased in energy by 1.13 eV relative to the S$_{0}$ minimum-energy geometry at the LR-TDDFT(PBE0)/aug-cc-pVDZ level of theory. In contrast, the S$_1$ ($\pi^{\ast}\leftarrow n$) state is stabilised, decreasing the energy gap with the S$_{0}$ state while increasing the gap with the higher-lying singlet states (S$_{n}$; $n > 1$). The predicted emission energy from the S$_{1}$ minimum-energy geometry is \textit{ca.} 3.9 eV, a value that is in excellent agreement with the emission spectrum recorded by Lee \textit{et al.}\cite{lee1971fluorescence} The broad band observed in the absorption spectrum recorded by Diau \textit{et al.}\cite{diau2001femtochemistry} is consistent with a short-lived electronically-excited state, however, which is indicative of competitive photochemical channels, \textit{e.g.} non-radiative decay on the femto/picosecond timescale through accessible CIs as discussed by Liu \textit{et al.}\cite{liu2016new}

The S$_{2}$ state minimum energy geometry (Table S4), in contrast, is reached from the Franck-Condon point \textit{via} contraction of the C=O bond (1.21 to 1.16 \mbox{\AA}) and slight elongation of the C-C bonds in the cyclobutane ring. The largest influence on energy observed is a destabilisation of the ground state, which increases in energy by 0.51 eV with a similar (0.58 eV) increase observed for the S$_1$ state. The vibronic structure observed in the absorption spectrum for this state exhibits a distinct vibronic structure \cite{diau2001femtochemistry} indicating both a longer lived excitation state and dominant vibrational modes activated upon excitation, which will be discussed in the following section. 

Previous studies, \textit{e.g.} Kao \textit{et al.},\cite{kao2020effects} have hypothesised as to the potential role of triplet states in the photochemistry of cyclobutanone; this is our motivation for presenting these states (T$_{n}$; $n = [1..5]$) in Table \ref{tab:es} and the relevant state-to-state spin-orbit couplings (SOCs) in the SI (Table S7). At the Franck-Condon geometry, the lowest-energy electronically-excited triplet state (T$_1$) is located at 3.73 eV above the electronic ground state at the LR-TDDFT(PBE0)/aug-cc-pVDZ level of theory and is of the same $\pi^{\ast}\leftarrow n$ character as the S$_1$ state; consequently, S$_{1}$/T$_{1}$ SOC will be formally forbidden.\cite{penfold2010effect,penfold2018spin} The T$_{2}$ and T$_{3}$ states are near-degenerate and located a little under the S$_{2}$ (6.10 eV) at 5.98 and 6.02 eV, respectively, above the electronic ground state at the LR-TDDFT(PBE0)/aug-cc-pVDZ level of theory. The SOC between these triplet states and the low-lying singlet electronically-excited states is generally small (Table S7), suggesting that the triplet states are unlikely to play a significant role in the early-time (\textit{e.g.} $<$2 ps) photochemistry focused upon in the present work. 

\begin{figure}[ht]
    \centering
    \includegraphics[width=0.95\linewidth]{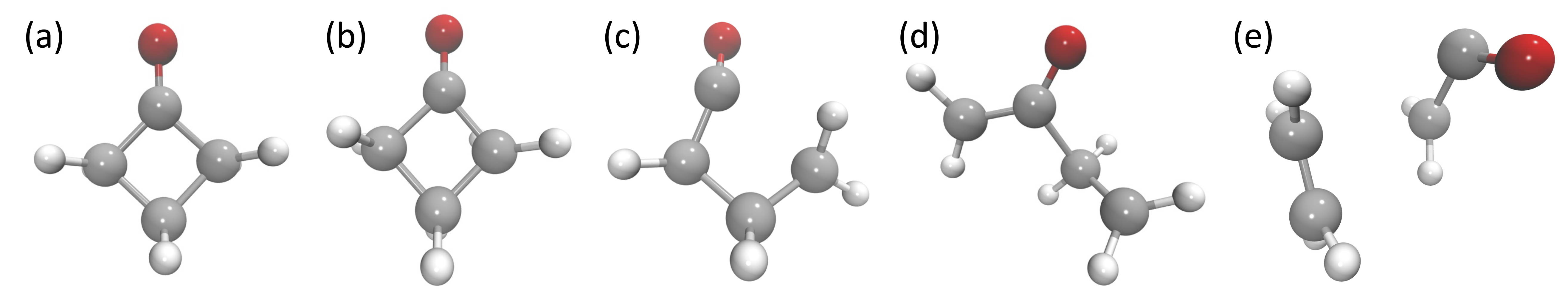}
    \caption{Key geometries of cyclobutanone: (a) the S$_{0}$ minimum-energy geometry, (b) the (symmetry-broken) S$_2$/S$_1$ MECI, and the S$_1$/S$_0$ (c) C$_{\alpha}$-cleavage, (d) C$_{\beta}$-cleavage, and (e) concerted C$_{\alpha}$-/C$_{\beta}$-cleavage MECIs. All geometries were optimised at the LR-TDDFT(PBE0)/aug-cc-pVDZ level of theory.}
    \label{fig:keygeoms}
\end{figure}

Figure \ref{fig:keygeoms}b shows the structure of an S$_{2}$/S$_{1}$ MECI, while Figures \ref{fig:keygeoms}c-e show the structures of three S$_{1}$/S$_{0}$ MECI. Cartesian coordinates are given in the SI. Potential energy surface(s) between the Franck-Condon point and each of the S$_{1}$/S$_{0}$ MECI were calculated \textit{via} linear interpolation in internal coordinates (LIIC) at the LR-TDDFT(PBE0) level are also given in the SI. 

The S$_{2}$/S$_{1}$ MECI (Figure \ref{fig:keygeoms}b) is located at 5.85 eV above the S$_{0}$ minimum-energy geometry, \textit{i.e.} \textit{ca.} 0.5 eV below the $3s\leftarrow n$ excitation energy, at the LR-TDDFT(PBE0)/aug-cc-pVDZ level of theory. Its structure is similar to the Franck-Condon geometry although it features a symmetry-breaking distortion of the cyclobutane ring (seen clearly on inspection of the position of the hydrogen atoms in Figure \ref{fig:keygeoms}b). The potential energy surface between the Franck-Condon geometry and this S$_{2}$/S$_{1}$ MECI is barrierless (Figure S3) which would suggest that S$_{2}$/S$_{1}$ internal conversion should be (ultra)fast in the absence of any additional considerations, however the symmetry of the two states at the Franck-Condon geometry is such that the interstate coupling is forbidden; even at the distorted S$_{2}$/S$_{1}$ MECI geometry, it is weak and results in a predictably longer lifetime for the S$_{2}$ state.

The three S$_{1}$/S$_{0}$ MECI (Figure \ref{fig:keygeoms}c-e) obtained are in close qualitative agreement (geometrically and energetically) with those located at the CASPT2 level of theory and reported in Ref. \cite{liu2016new}. The first is located along the $\alpha$-cleavage channel, the second along the $\beta$-cleavage channel, and the third along a concerted $\alpha$/$\beta$-cleavage channel. The energetic ordering of the three S$_{1}$/S$_{0}$ MECIs (3.40, 2.56 and 5.0 eV, respectively, above the S$_{0}$ minimum-energy geometry at the LR-TDDFT(PBE0)/aug-cc-pVDZ level of theory) follows qualitatively the trend observed for the three S$_{1}$/S$_{0}$ MECIs in Ref. \cite{liu2016new}, although it is important to note that the single-reference nature of LR-TDDFT(PBE0) favours charged rather than biradical dissociation along the $\alpha$- and $\beta$-cleavage channels and, furthermore, renders it unable to describe properly the topology/dimensionality of the S$_{1}$/S$_{0}$ crossing seam. Although all of the S$_{1}$/S$_{0}$ MECIs are energetically accessible (\textit{i.e.} they and their barriers are submerged relative to the \textit{ca.} 6.2 eV excitation energy, the description of the potential energy surface around the S$_{1}$/S$_{0}$ MECIs is likely to be problematic for LR-TDDFT(PBE0) and it is quite possible that this might influence the photoproduct production by influencing the internal conversion dynamics through the MECI and, subsequently, the motion of the trajectory/wavepacket on the electronic ground state potential energy surface. 

Overcoming these aforementioned limitations could be achieved using a multireference (active space) method, \textit{e.g.} CASSCF/CASPT2 or NEVPT2, for the excited-state molecular dynamics simulations. However, the performance of these methods is greatly dependent on the choice of active space; an appropriate active space should be large enough to incorporate all of the orbitals required over all of the nuclear configurations explored in the excited-state molecular dynamics simulations while not too large so as to render the simulations computationally costly to the point of intractability. We found an active space of eight electrons in eight orbitals [\textit{e.g.} NEVPT2(8,8)] unstable with respect to orbital rotation(s) at some distorted cyclobutane geometries, while a larger active space of twelve electrons in twelve orbitals [\textit{e.g.} NEVPT2(12,12)] was too computationally expensive to carry out practicably excited-state molecular dynamics simulations. Consequently, in the present work, we have carried out our excited-state molecular dynamics simulations using LR-TDDFT(PBE0) and ADC(2), keeping in mind the aforementioned (although well-understood) limitations and their potential impact on the dynamics (which we discuss in detail). However, we note within the context of the present challenge, other contributors have performed excited-state molecular dynamics simulations with CASSCF based on an eight-electron-in-eleven-orbital active space\cite{daria2024photofragmentation} and extended multistate CASPT2 (XMS-CASPT2) based on an eight-electron-in-eight-orbital active space.\cite{janos2024predicting}

\subsection{Early-Time Dynamics Using a Spin-Vibronic Coupling Hamiltonian}

To i) identify possible photochemical (fragmentation) channels, ii) clarify the potential involvement of triplet states, and iii) establish a mechanism for internal conversion from the initially-excited S$_2$ ($3s \leftarrow n$) state to the lowest-energy singlet electronically-excited state (S$_{1}$) at early times, we employ a model Hamiltonian and carry out quantum dynamics simulations as described above. The model Hamiltonian comprises the electronic ground state (S$_{0}$) and nine electronically-excited states [four singlets (S$_{n}$; $n = [1..4]$) and five triplets (T$_{n}$; $n = [1..5]$)]. The inclusion of singlet states higher in energy than the S$_2$ (\textit{e.g.} S$_{n}$; $n > 2$) is motivated by the absence of vibronic coupling between the S$_{1}$ and S$_{2}$ states, both of which are of A" symmetry (Table \ref{tab:es}); here, coupling to higher-lying singlet states of, \textit{e.g.}, alternative symmetries offers the potential of second-order population transfer channels comparable to those identified in other systems where direct population transfer channels are weak or otherwise absent.\cite{CMA_Eng2018,CMA_Eng2020,CMA_Eng2022} 

\begin{figure}[ht]
    \includegraphics[width=0.75\linewidth]{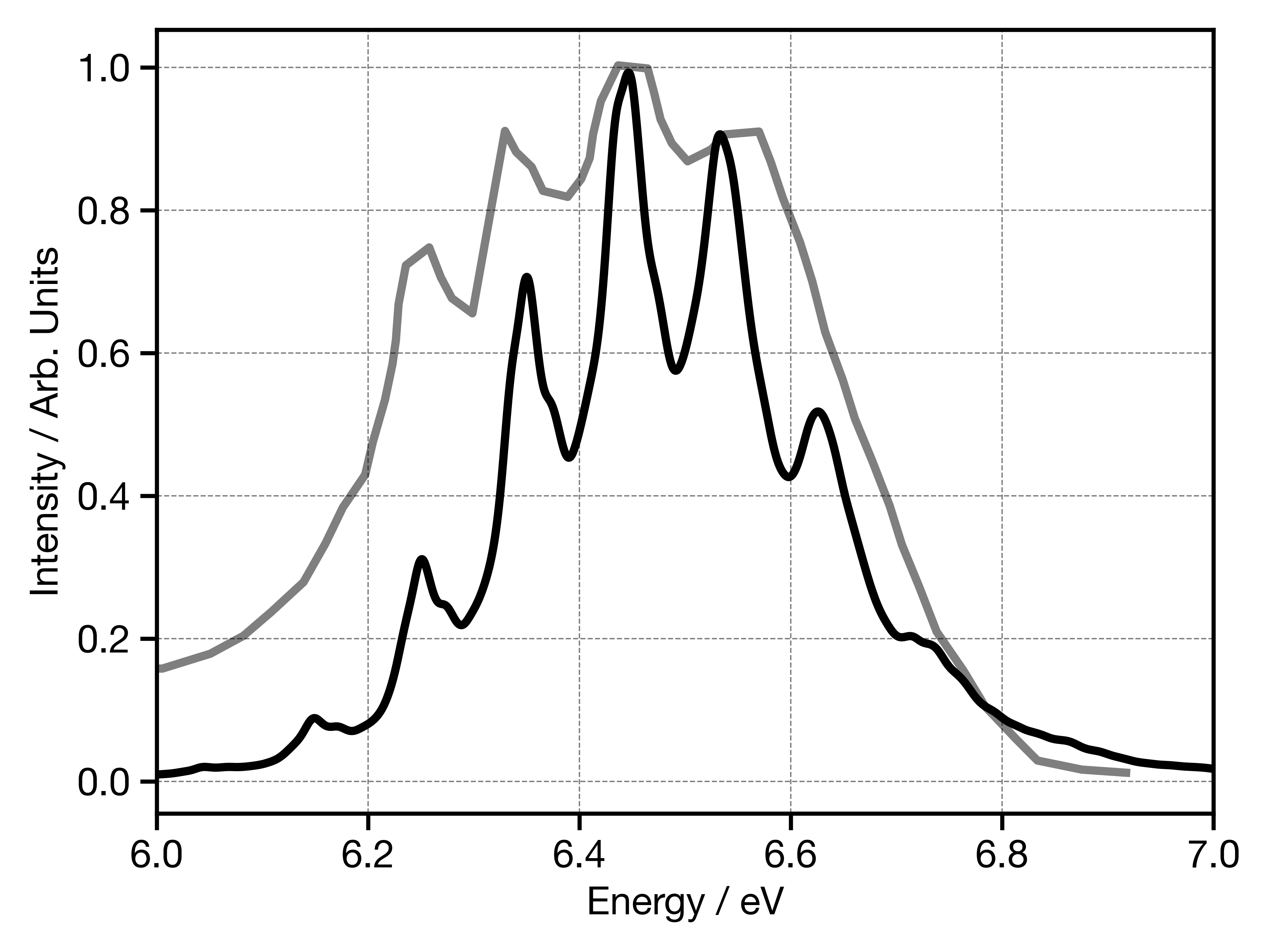} 
    \caption{Theoretical (black) and experimental (grey) S$_{2}$ $\leftarrow$ S$_{0}$ ($3s \leftarrow n$) absorption spectrum. The theoretical spectrum is shown shifted by $\Delta{E} = +0.14$ eV. The experimental spectrum was recorded by Diau \textit{et al.} and digitised from Ref. \cite{diau2001femtochemistry}}
    \label{fig:abs}
\end{figure}

In nuclear configurational space, the model Hamiltonian incorporates eight degrees of vibrational freedom: $\nu_{1}$, $\nu_{7}$, $\nu_{10}$, $\nu_{11}$, $\nu_{12}$, $\nu_{13}$, $\nu_{15}$, and $\nu_{21}$, which were identified by the magnitude of their first-order couplings and the symmetries of the vibrational modes. The degrees of vibrational freedom included the model Hamiltonian comprise cyclobutane ring puckering ($\nu_{1}$ and $\nu_{12}$), symmetric and anti-symmetric cyclobutane ring breathing ($\nu_{7}$ and $\nu_{10}$), cyclobutane ring deformation ($\nu_{11}$, $\nu_{13}$ and $\nu_{15}$), and the C=O stretching mode ($\nu_{21}$), consistent with previous works.\cite{kuhlman2012coherent} To assess the approximate accuracy of our model Hamiltonian, Figure \ref{fig:abs} compares the experimental \cite{diau2001femtochemistry} and theoretical absorption spectrum for the S$_2$ state. There is excellent agreement between the two, indicating that our model Hamiltonian describes accuracy the (local) potential energy surface(s). The absorption spectrum features a distinct vibrational progression with structured peaks separated by \textit{ca.} 0.1 eV consistent with the C=O out-of-plane wagging and cyclobutane ring puckering vibrational modes previously identified \cite{baba1984s} and arising in the present model \textit{via} the inclusion of $\nu_{12}$.

\begin{figure}[ht]
    \centering
    \includegraphics[width=0.7\linewidth]{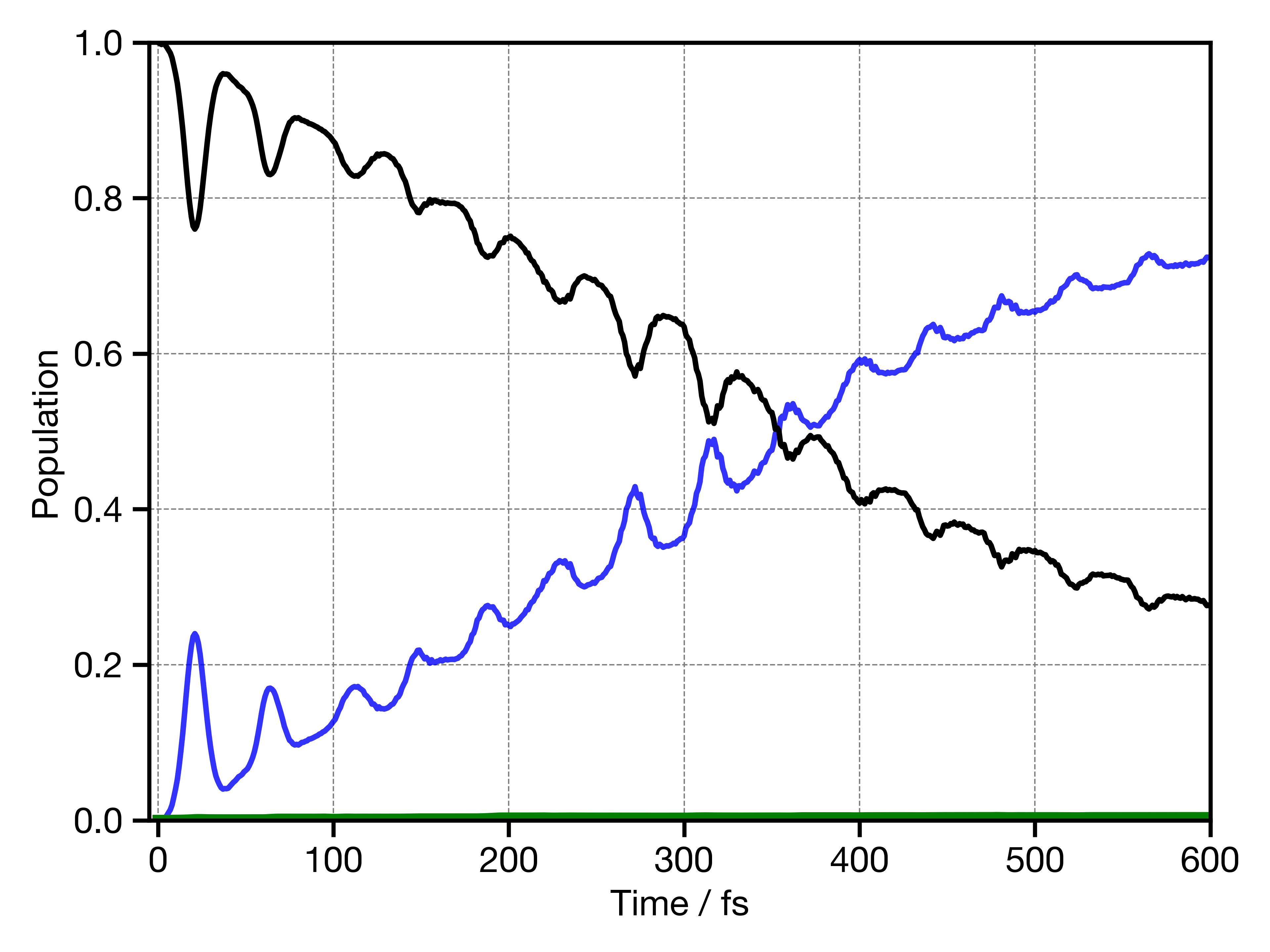}\\
    \caption{Population kinetics obtained from quantum dynamics simulations over 600 fs post-photoexcitation into the S$_{2}$ ($3s \leftarrow n$) state. The S$_{2}$ state population is shown in black; the S$_{1}$ state population is shown in blue; the triplet (T$_{n}$; $n = [1..5]$) state populations are shown (collectively) in green.}
    \label{fig:qd}
\end{figure}

Figure \ref{fig:qd} shows the transfer of population from the S$_2$ state over 600 fs post-photoexcitation into the S$_2$ ($3s \leftarrow n$) state as computed \textit{via} quantum dynamics. The S$_{1}$ $\leftarrow$ S$_{2}$ population transfer occurs at a rate of \textit{ca.} $1.67\times10^{-12}$ s$^{-1}$ with the S$_{1}$ state reaching peak population within \textit{ca.} 350 fs and is qualitatively consistent with the decay of the peak associated with the S$_{2}$ state in the photoelectron spectrum recorded by Kuhlman \textit{et al.}\cite{kuhlman2012coherent} (the authors report a biexpotential fit yielding time constants of \textit{ca.} 350 and 750 fs). The S$_{1}$ $\leftarrow$ S$_{2}$ population transfer in the model Hamiltonian obtains intensity \textit{via} mixing with the higher-lying singlet electronically-excited states, especially the S$_4$ state (the lowest-lying singlet electronically-excited state of A' character). This occurs most prominently along $\nu_{11}$. While this mode is responsible for the state-to-state coupling, it is $\nu_{1}$, $\nu_{12}$, $\nu_{13}$, and $\nu_{21}$ which exhibit the largest-amplitude nuclear oscillations and electronically-excited-state structural changes which are likely to be observed in the ultrafast electron diffraction experiment. The population oscillations observed between the S$_{2}$ and S$_{1}$ states are associated with the overcoherence of the reduced model Hamiltonian and are, in any case, faster than the temporal resolution of the ultrafast electron diffraction experiments.

S$_{0}$ $\leftarrow$ S$_{1}$ population transfer is not included in the present model Hamiltonian. The high energy of the S$_2$ state results in an exceptionally hot wavefunction on the electronic ground state that is difficult to converge under the framework of the quantum dynamics simulations. In addition, previous works -- and our own quantum chemical calculations -- have identified S$_{1}$/S$_{0}$ MECI at highly distorted geometries beyond the limits of the normal model representation on which the quantum dynamics simulations are predicated (the representation is only valid to small distortions from the equilibrium, \textit{e.g.} S$_{0}$ minimum-energy, geometry). Consequently, to describe more completely the excited-state relaxation dynamics, we explore excited-state molecular dynamics simulations operating in unconstrained nuclear configurational space through the trajectory surface-hopping approach. 

\subsection{Excited-State Relaxation Dynamics in Unconstrained Nuclear Configuration Space Using Trajectory Surface-Hopping Dynamics}

\begin{figure}[ht]
    \centering
    \includegraphics[width=0.75\linewidth]{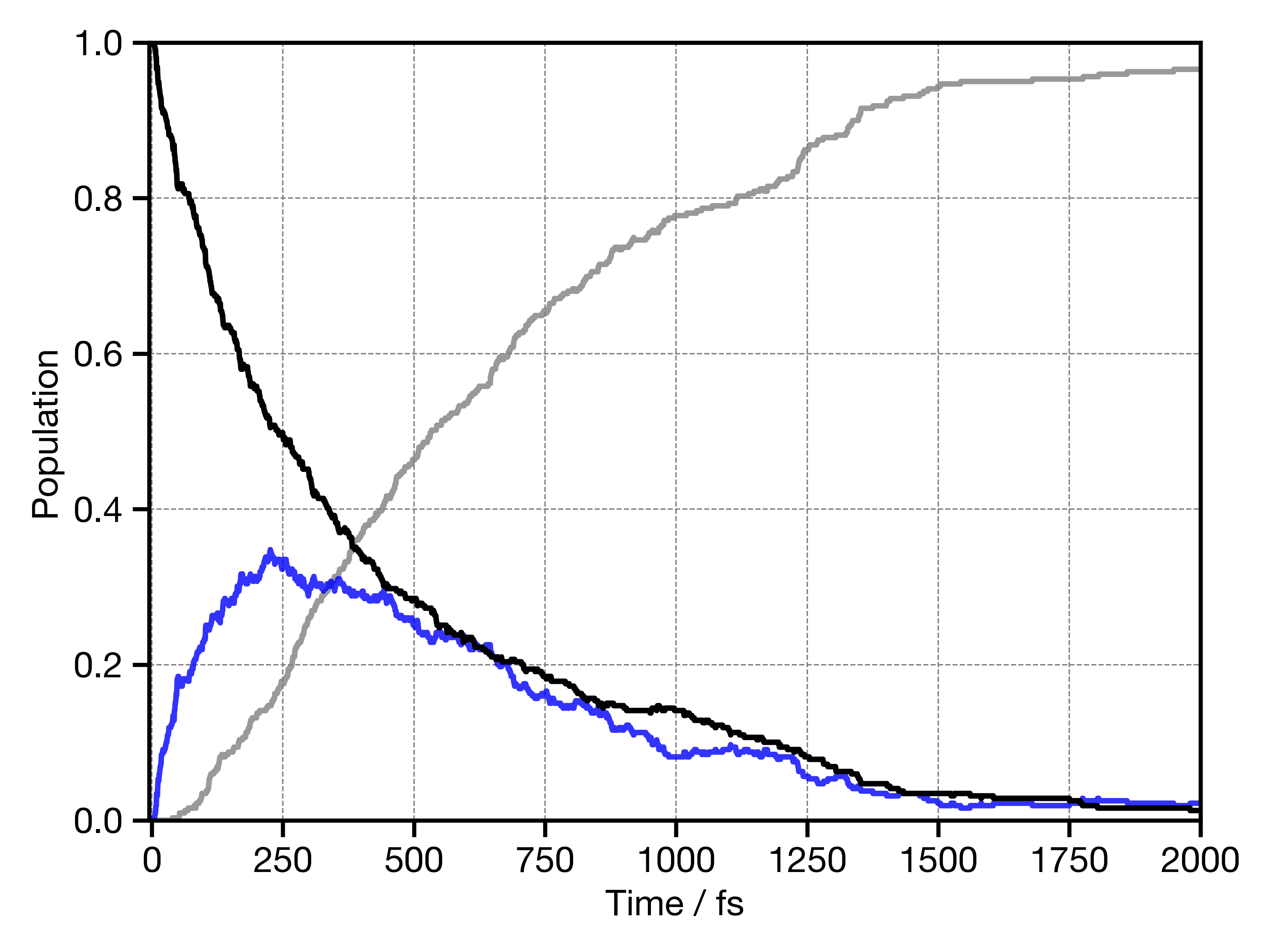}\\
    \caption{Population kinetics obtained from 289 \textit{on-the-fly} surface-hopping trajectories over 2 ps post-photoexcitation into the S$_{2}$ ($3s \leftarrow n$) state. The S$_{2}$ state population is shown in black; the S$_{1}$ state population is shown in blue; the S$_{0}$ state population is shown in gray.}
    \label{fig:TSHpopkinetics}
\end{figure}

Figure \ref{fig:TSHpopkinetics} shows the populations of the S$_{2}$, S$_{1}$ and S$_{0}$ states over the first 2 ps post-photoexcitation into the S$_2$ ($3s \leftarrow n$) state as obtained from 289 \textit{on-the-fly} surface-hopping trajectories propagated at LR-TDDFT(PBE0)/aug-cc-pVDZ level of theory. A similar set of trajectory surface-hopping dynamics propagated at the ADC(2)/aug-cc-pVDZ level of theory are presented and analysed in the SI and are of comparative interest. 

Figure \ref{fig:TSHpopkinetics} shows S$_{1}$ $\leftarrow$ S$_{2}$ population transfer with a decay constant of \textit{ca.} 356 fs. This is slightly faster than observed in the quantum dynamics simulations, \textit{i.e.} 50\% of population decays from S$_2$ in 225 fs, which is due to the inclusion of the ground state to which the wavepacket can rapidly decay. This decay constant is in close agreement with the fastest time-constant extracted from a photoelectron spectroscopic study,\cite{kuhlman2012coherent}, but we do not see any dynamics associated with the $\sim$700 fs component reported. Interestingly, this slower component is in close agreement with the population kinetics for the dynamics performed using potentials calculated at ADC(2) level of theory, shown in the supporting information.

\begin{figure}[ht]
    \centering
    \includegraphics[width=0.75\linewidth]{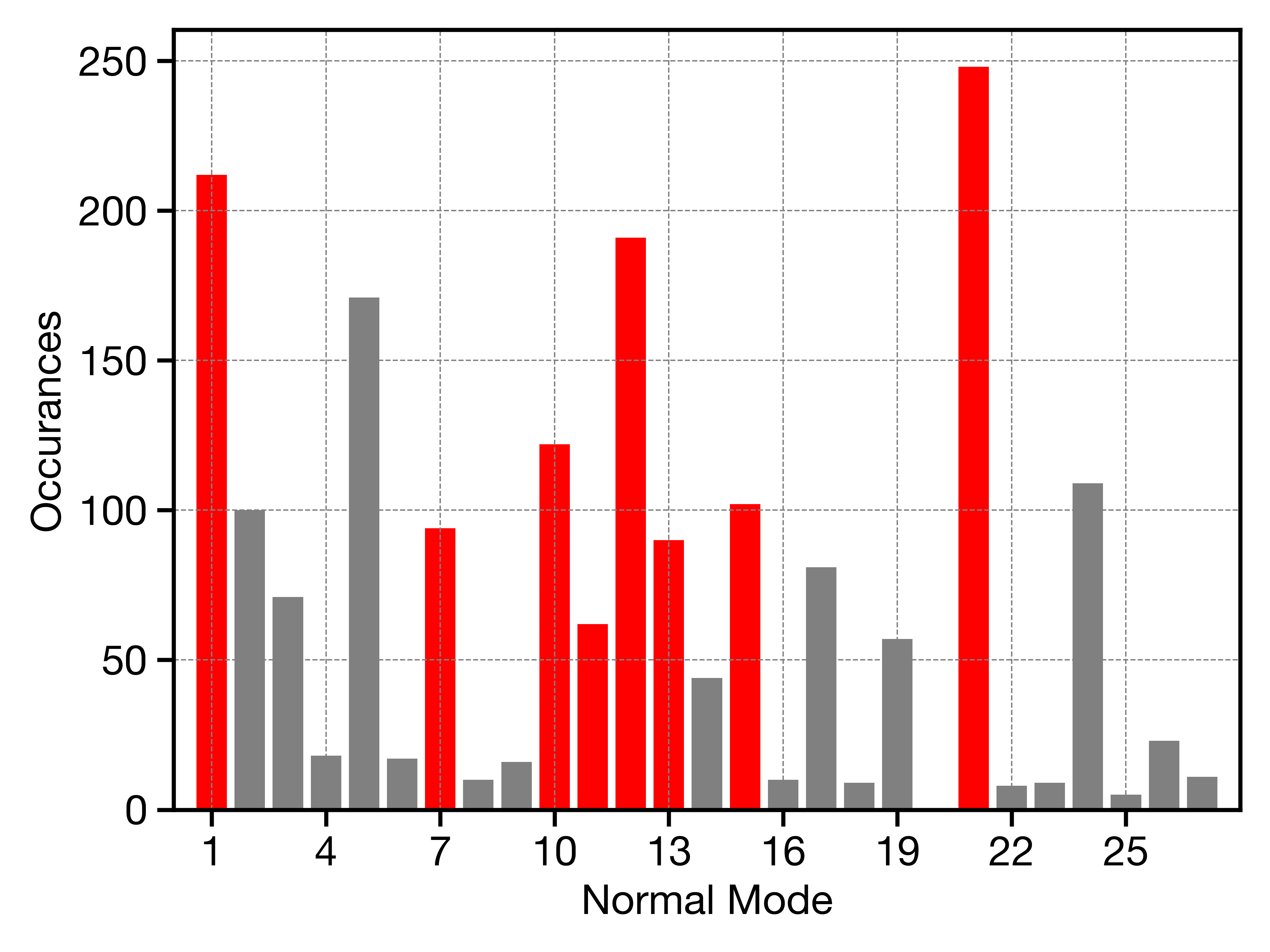}
    \caption{Bar chart showing the the number of times that each normal mode features in the top six largest displacements for an (electronically-excited-state) trajectory. The red bars represent those normal modes included in the model Hamiltonian.}
    \label{fig:histotshqd}
\end{figure}

While the time scales between the quantum dynamics and TSH simulations suggests similar dynamics, further analysis is required to assess this in more detail. To achieve this, we transform the first 500 fs of excited state molecular dynamics from Cartesian coordinates into a normal mode representation similar to ref. \cite{capano2017photophysics}. Figure \ref{fig:histotshqd} shows the normal modes active during the TSH simulations as a bar chart of the number of times each normal mode features in the top 6 of the largest displacements during a trajectories excited state dynamics. The red bars represent those normal modes included in the model Hamiltonian. This shows close agreement between the modes which appear most frequently in the TSH simulations and those included in the quantum dynamics. The most notable exception is $\nu_5$ which exhibits larger amplitude motion due to the flat nature of the potential, but does not act as either tuning or coupling modes and therefore do not strongly influence the excited state dynamics.

\begin{figure}[ht]
    \centering
    \includegraphics[width=0.9\linewidth]{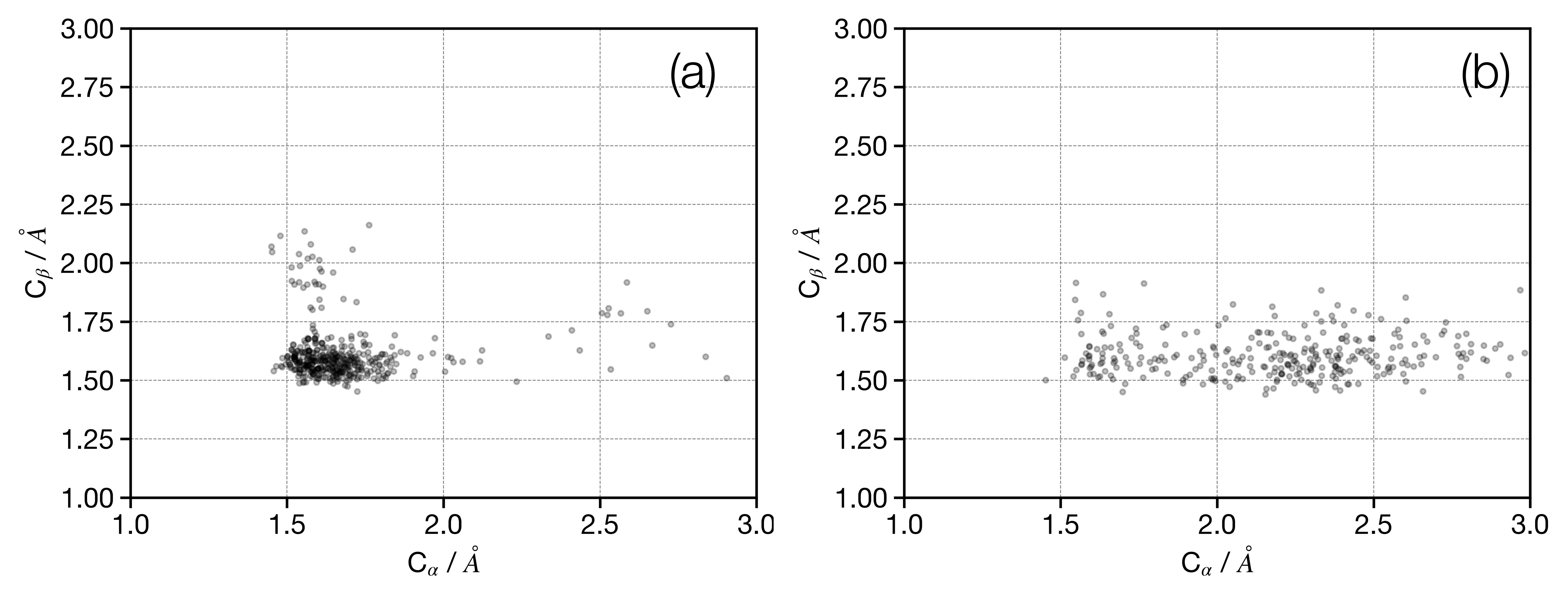}
    \caption{Average C$_{\alpha}$ and C$_{\beta}$ bond lengths at (a) S$_{1}$ $\leftarrow$ S$_{2}$ and (b) S$_{0}$ $\leftarrow$ S$_{1}$ surface-hopping events.}
    \label{fig:tdstructuralparms}
\end{figure}

Figure \ref{fig:tdstructuralparms} shows average C$_{\alpha}$ and C$_{\beta}$ bond lengths for the structures where each trajectory hops from the S$_2$-S$_1$ (a) and S$_1$-S$_0$ (b) states. For the former, there is a clear cluster around 1.5-1.6 \mbox{\AA} consistent with the ground state structure and therefore close to the Franck-Condon geometry, as expected from the optimised  S$_2$/S$_1$ discussed above. In contrast, for the S$_1$-S$_0$ hopping geometries, there is a significant change, with the majority of \textit{hops} occurring for C$_{\alpha}$ bond lengths $>$2.2 \mbox{\AA}. Using the geometries provided in the supporting information, the C$_{\alpha}$ CI occurs when the C$_{\alpha}$ bond length is 2.35 \mbox{\AA}, with very little correspond change along the C$_{\beta}$ bond. This suggests, in agreement with previous work, that crossing from the S$_1$-S$_0$ occurs primarily at the CI exhibiting a C$_{\alpha}$ bond break.

\begin{figure}[ht]
    \centering
    \includegraphics[width=0.9\linewidth]{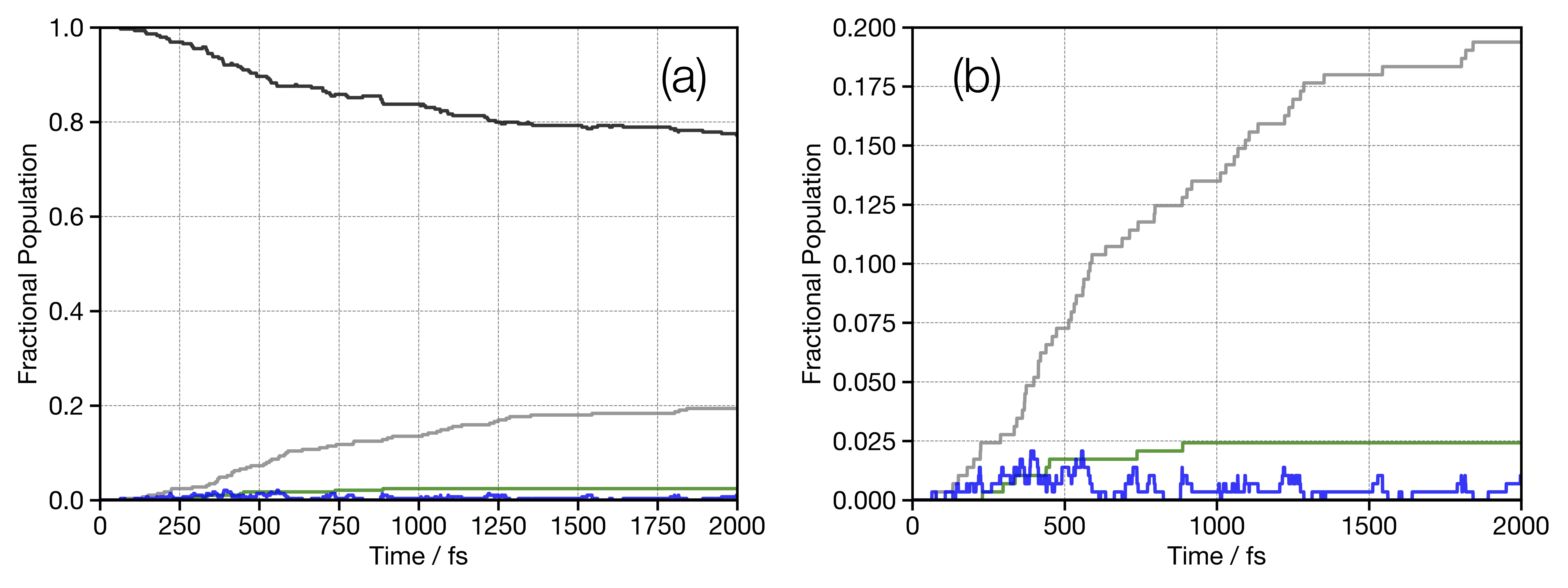}
    \caption{Fractional population of the photoproducts obtained from the 289 trajectories. The black trace shows cyclobutanone, the grey trace shows the \textbf{C2} products (C$_2$H$_4$ + CH$_2$CO), the green trace shows the \textbf{C3} products (C$_3$H$_6$ + CO), and the blue trace shows ring-opened structures.}
    \label{fig:photoproducts_pop}
\end{figure}

Figure \ref{fig:photoproducts_pop} shows the fractional population of the photoproducts formed from the TSH trajectories. This indicates $\sim$20\% of the excited states form the C$_2$H$_4$ + CH$_2$CO, \textit{i.e.} decay \textit{via} the \textbf{C2} channel, while 2.5\% forms the \textbf{C3} products. The ring-open species are formed, but are very short-lived and either contribute forming either the \textbf{C2} or \textbf{C3} products or undergo bond reformation to form vibrationally excited ground state cyclobutanone. The formation of \textbf{C2} comparable, but lower than other excited state dynamics simulations performed at a higher level of theory reported in ref. \cite{janos2024predicting} (34\%) and with Trentelman \textit{et al.} \cite{trentelman1990193} who reported 43\% of yield experimentally. The major discrepancy in our simulations occurs for the \textbf{C3} channels, which is $>$60\% in these previous works. The near-absence of the \textbf{C3} channel is associated with the multi-reference character of the potential in this region and the bias of single reference methods for charged rather than biradical bond breaking. While this does not significantly increase the energies of the C$_{\alpha}$ and C$_{\beta}$ CIs (see Figure S4), it does increase the energy of the double bond breaking CI, making the formation of the \textbf{C3} channel challenging. To assess this we also perform dynamics using trajectories in the T$_1$ state, performed using unrestricted Kohn-Sham facilitating the description of biradical character. These were initiated at random from trajectories populating the S$_1$ state. Importantly, these show a much higher formation of \textbf{C3} photoproducts (\textbf{C3}: 53\%, \textbf{C2}: 5\% and ring-open: 17\%) consistent with previous experiments \cite{trentelman1990193}.

\begin{figure}[ht]
    \centering
    \includegraphics[width=0.9\linewidth]{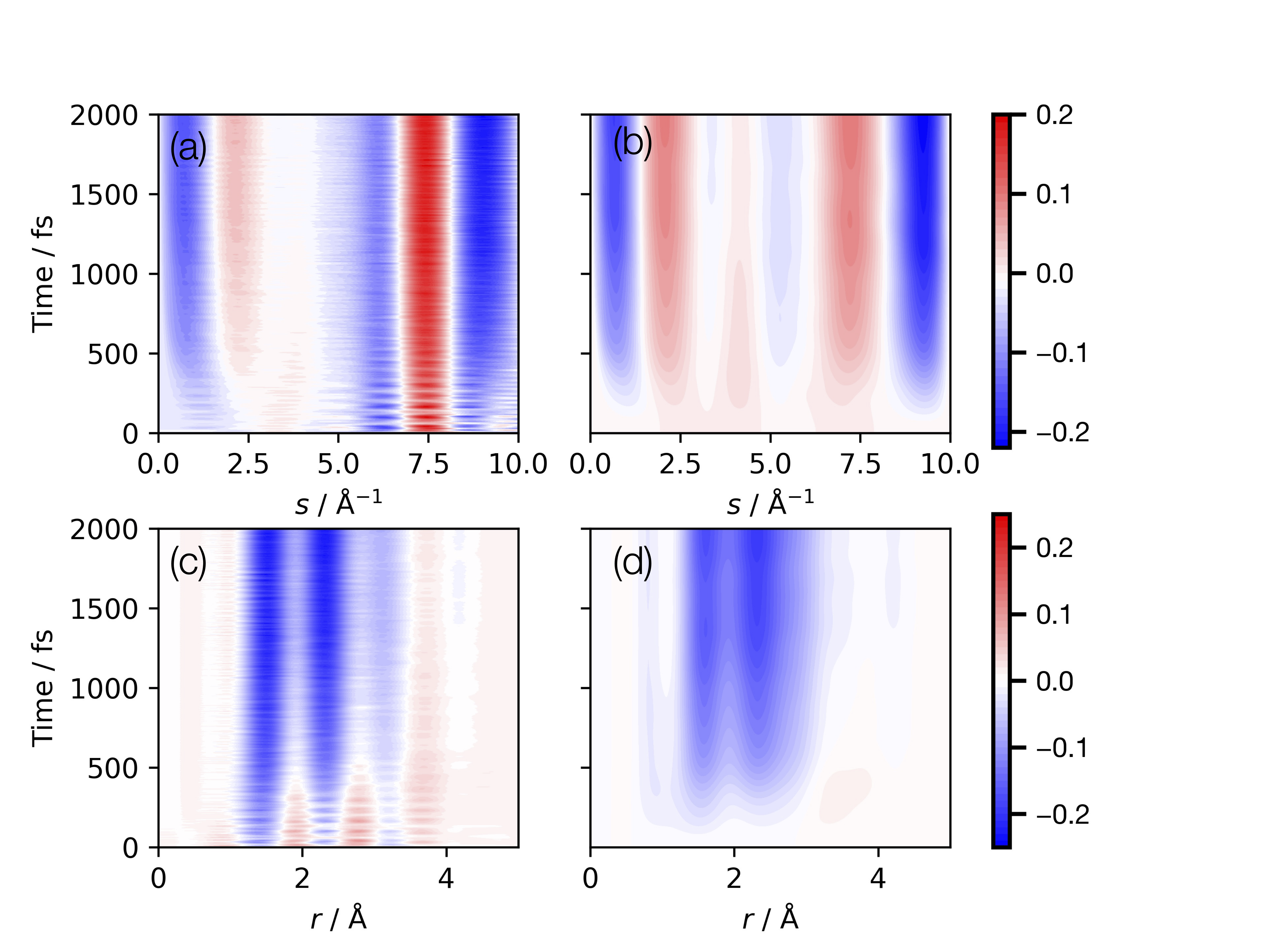}\\
    \caption{Transient ($\Delta I/I$) scattering (a) without and (b) with 150 fs (FWHM) temporal broadening. Transient PDF (c) without and (d) with 150 fs (FWHM) temporal broadening. The ground-state (pre-photoexcitation) signal used to generate the transient signal was obtained from the trajectory surface-hopping dynamics initial conditions, \textit{i.e.} the nuclear ensemble representing the state of the system at $t = 0$. All plots were produced using the 289 2000-fs trajectories simulated at the LR-TDDFT(PBE0)/aug-cc-pVDZ level of theory.}
    \label{fig:trelectrondiffraction}
\end{figure}

\subsection{Electron Diffraction Simulations}

Figure \ref{fig:trelectrondiffraction} shows the time resolved electron diffraction simulations arising after photoexcitation of cyclobutanone into the S$_2$ state. Figure \ref{fig:trelectrondiffraction}a show the electron diffraction scattering signal as calculated, while Figure \ref{fig:trelectrondiffraction}b is convolved along the temporal axis with a Gaussian kernel (FWHM = 150 fs) to reproduce the effect(s) of the finite temporal resolution of the proposed electron diffraction experiment. Figures \ref{fig:trelectrondiffraction}c and d show time-resolved pair-distribution function (PDFs) maps with and without temporal broadening, respectively, and were produced \textit{via} sine transformation of the modified scattering intensity maps in Figures \ref{fig:trelectrondiffraction}a and b, respectively.

The modified scattering intensity maps (Figures \ref{fig:trelectrondiffraction}a and b) show two strong negative (\textit{ca.} 1 and 9 \mbox{\AA}$^{-1}$) and two positive (\textit{ca.} 2.7 and 7.5 \mbox{\AA}$^{-1}$) features but do not reveal the richness of the dynamics that reflect the complex photochemistry of cyclobutanone, in part due to the incoherent/stochastic nature of the photochemical processes taking place.

\begin{figure}[ht]
    \centering
    \includegraphics[width=0.7\linewidth]{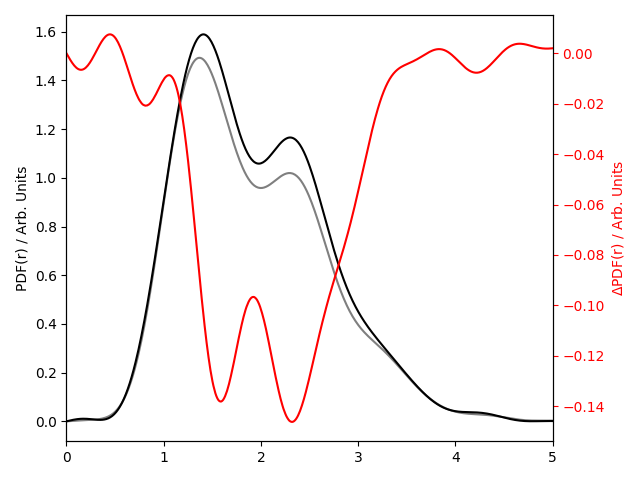}\\
    \caption{Initial ($t = 0$ fs; black) and final ($t = 2000$ fs; grey) PDFs and the difference PDF (red) calculated using the 289 2000 fs surface hopping trajectories simulated using potentials are LR-TDDFT(PBE0) level of theory.}
    \label{fig:pdf}
\end{figure}

A deeper understanding of the structural changes can be established from Figures \ref{fig:trelectrondiffraction}c and d, which show the time-resolved PDF. For clarity, Figure \ref{fig:pdf} shows the PDF($t = 0$ fs), PDF($t = 2000$ fs), and the $\Delta$PDF($t = 2000$ fs).The  PDF acquired for the 289 initial conditions exhibits 3 peaks at $\sim$1.5 \mbox{\AA}, $\sim$2.5 \mbox{\AA} and $\sim$3.0 \mbox{\AA}. The first corresponds to the neighing C-C and C=O distances, the second corresponds to C-C distance on the opposite side of the cycle and the final peak corresponds to the C-O distances which are not directly bonded. The transient indicates primarily a loss of the first two peaks associated with dissociation and the formation of the \textbf{C2} products.  We note here that despite the aforementioned differences in the photoproduct formation, the transient scattering and PDFs are very similar to those in ref. \cite{janos2024predicting}. This highlights the challenge in disentangling the exact photoproduct formation of these systems due to the overlapping bands.

\section{Discussion and Conclusions}

In this work, we have carried out quantum and excited-state trajectory surface-hopping molecular dynamics simulations to study the electronically-excited-state relaxation mechanisms and electronic ground-state dynamics of cyclobutanone post-photoexcitation into the S$_{2}$ Rydberg ($3s \leftarrow n$) state. Our focus has been upon translating these simulations to predict the experimental observables associated with the ultrafast electron diffraction experiments and ultimately to answer the question: \textit{Can excited state dynamics simulations be predictive?} However, even for small molecules such as cyclobutanone, certain approximations in the underlying computational chemical methods are required which will influence the outcome of such simulations: we highlight this in comparison between the present work and other works related to the same challenge.\cite{miao2024casscfmrci,suchan2024prediction,daria2024photofragmentation,janos2024predicting} Consequently, in this section we discuss the relaxation mechanism observed in our simulations as well as potential sources of error and how we expect that these will influence interpretation of the experimental observables. 

Within this challenge, the objective has been to translate excited state dynamics into experimental observables. The importance of this cannot be understated. In many cases, collaboration between experimental and theoretical studies focus upon the comparison of quantities that are easy to calculate, such as electronic state population kinetics. These kinetics are then compared to experimentally extracted timescales and agreement is taken as accuracy of the simulations. However, as shown in this work the LR-TDDFT (356 fs) and XMS-CASPT2 \cite{janos2024predicting} (335 fs) dynamics provide very similar decay kinetics, but different predictions of photoproducts which would influence the experimental signal.  While slightly slower, the excited state dynamics performed using ADC(2) potentials also occurs on a comparable timescale ($\sim$700 fs), but owing to the artificial crossing along the C=O bond stretch \cite{marsili2021caveat}, the excited state decay occurs via a completely different mechanism. 

Our simulations indicate that after excitation of the $3s \leftarrow n$ Rydberg state the system relaxes within 1-2 ps to form a broad range of photoproducts. Decay of this initially excited S$_2$ state occurs with a time-constant of $\sim$350 fs. This is in good agreement with the fastest kinetics reported from previous time-resolved photoelectron experiments by Kuhlman \textit{et al.} \cite{kuhlman2012coherent}. However, we note that ref. \cite{kuhlman2012coherent} also reports a strong contribution from a slower time component, $\sim$750 fs. This is not observed within the LR-TDDFT population kinetics, but is in very close agreement with the ADC(2) kinetics presented in the supporting information. The exact origin for this difference between LR-TDDFT and ADC(2) is unclear, however analysis of the \textit{hopping} geometries, indicates a flatter potential in the case of the latter, which permits a slightly wider spread of the trajectories in nuclear configuration space. Importantly, in both cases despite the small nuclear displacement required to reach the crossing point, the internal conversion from S$_2$ to S$_1$ is comparatively slow due to the symmetry forbidden nature of the transition. 

Once populated, the S$_1$ undergoes a large structural distortion, primarily along the C$_{\alpha}$ bond, consistent with previous work \cite{diau2001femtochemistry,liu2016new}. This drives population to be rapidly transferred into the electronic ground state to form very vibrationally hot species. The fast nature of the population transfer from the S$_1$ to the ground state means that population of the S$_1$ doesn't exceed $\sim$30\%. The excited state molecular dynamics consider only the dynamics within the singlet manifold. To assess the potential influence of the intersystem crossing and the triplet states, our quantum dynamics include the low lying excited triplet states. These simulations indicate negligible amount of intersystem crossing into the triplet manifold leading us to conclude that this channel will be unable to compete with internal conversion rates found here. 

Although the excited state dynamics and high energy of excitation leads to some highly distorted geometries, our simulations point to the formation of the photoproducts being determined as the trajectory passes through the CI between the first electronically excited state and the ground state \textit{via} a ring-opened intermediate. This is consistent with the conclusions from ref. \cite{diau2001femtochemistry} who demonstrated that motion away from the CI branching space leads to all of the observed photoproducts. Herein lies the most significant approximation within our work, as neither of the single reference method used will capture the biradical nature of the photoproducts. Indeed, although all of the CI identified \cite{liu2016new} in previous work have been found within the LR-TDDFT(PBE0) framework and exist at accessible energies, \textit{i.e.} below the excitation energy, our simulations show a much lower fraction of photoproduct formation than previous theoretical \cite{miao2024casscfmrci,suchan2024prediction,daria2024photofragmentation,janos2024predicting} and experimental \cite{trentelman1990193} works. This appears to most strongly affect the \textbf{C3} photoproducts, which only forms in 2.5\% of the trajectories. In contrast, trajectories in the T$_1$ state, performed using unrestricted Kohn-Sham facilitating the description of biradical character, initiated at random from trajectories populating the S$_1$ state, show a  much higher formation of \textbf{C3} photoproducts (\textbf{C3}: 53\%, \textbf{C2}: 5\% and ring-open: 17\%) consistent with previous experiments \cite{trentelman1990193}. The motion through the CI and therefore the potential shape in this region is likely to be critical in determining the branching ratio of the photoproducts. Here it may not only be a limitation of single reference methods but a also condition of the excited state dynamics used. As stated in the methods section, our present dynamics attempts to avoid instabilities in the multi-configurational region, near the degeneracy by enforcing a hop to the ground state when the S$_1$-S$_0$ energy gap became $\Delta\mathrm{E_{S_1-S_0}}<0.1\mathrm{eV}$. While this avoids the explicit motion through the CI, the enforced earlier transition may also promote populated transfer closer the cyclobutanone structure encouraging reformation of vibrationally hot cyclobutanone, rather than the photoproducts.

From these excited state molecular dynamics simulations, the ultrafast electron diffraction observable shows distinct changes and by studying the time-resolved PDF, these are largely associated with a loss in intensity for interactions at 1.5 and 2.5 \mbox{\AA}, arising from dissociation. Despite the rich dynamics and the distinct changes observed, the time-resolved scattering curves show very little distinct dynamics largely associated with the incoherent nature of the dynamics and the comparatively low temporal resolution (150 fs).

Finally, a logical question would be ask if the limitations discussed above can be overcome within the present framework, \textit{i.e.} without adopting a multi-reference wavefunction method which would prove challenging for larger systems. Here, it is important to stress that while the relative yields of photoproducts formed appears somewhat at odds with previous works, all major reported products are generated, \textit{i.e.} the full nuclear configuration space has been sampled. Excited state simulations have previously been used to simulate the experimental observables associated with structurally sensitive techniques of electron \cite{figueira2023monitoring} and X-ray diffraction \cite{minitti2015imaging}.  Importantly in both of these works the outcomes of the trajectory-based dynamics were used as a basis to fit to experimental data and deliver an interpretation. While, the use of a fit means that such an approach may not be classed as fully predictive, in both cases an excellent agreement between experiment and theory was achieved and providing deep insight into the dynamics observed.

\section*{Acknowledgements}

This research made use of the Rocket High Performance Computing service at Newcastle University and the Viking High Performance Computing service at the University of York. We also acknowledge the COSMOS Programme Grant (EP/X026973/1). T. J. P would like to thank the EPSRC for an Open Fellowship (EP/W008009/1).

\end{document}


\normalsize
  \begin{center}
\textbf{\Large{The Photochemistry of Rydberg Excited Cyclobutanone: Photoinduced Processes and Ground State Dynamics}}\\
\Large{Supplementary Information}

\vspace{0.5cm}
\noindent\large{J. Eng$^a$, C. D.  Rankine$^b$, T. J. Penfold$^a$} \\ \vspace{0.5cm}
{$^a$ Chemistry, School of Natural and Environmental Sciences, Newcastle University, Newcastle upon Tyne, NE1 7RU, UK}\\ \vspace{0.5cm}
{$^b$ Department of Chemistry, University of York, York, YO10 5DD, UK}\\ \vspace{0.5cm}
 {\textit{tom.penfold@newcastle.ac.uk}}\\
 {\textit{julien.eng@newcastle.ac.uk}}

 \end{center}

\tableofcontents
\listoftables
\listoffigures

\newpage
\pagestyle{fancy}
\fancyhead{} 
\fancyhead[L]{\leftmark}
\fancyhead[R]{\thepage \, of \pageref{LastPage}}
\fancyfoot{} 


\renewcommand*\rmdefault{bch}\normalfont\upshape
\rmfamily
\section*{}
\vspace{-1cm}


\footnotetext{\dag~Electronic Supplementary Information (ESI) available: [details of any supplementary information available should be included here]. See DOI: 10.1039/cXCP00000x/}


%
%

\section{Electronic Structure of Cyclobutantone}
\subsection{Method and Basis Set Dependence of Excited States at the Franck-Condon Geometry}

\begin{table}[h]
    \centering
    \begin{tabular}{|c|c|c|c|c|c|c|c|}
     \multicolumn{7}{c}{ADC(2)}\\
    \hline
    State &  cc-pVDZ & cc-pVTZ & cc-pVQZ & aug-cc-pVDZ & aug-cc-pVTZ & aug-cc-pVQZ \\ 
    \hline
      S$_1$    & 4.30 & 4.26  & 4.26 & 4.20 & 4.21 & 4.23 \\
      S$_2$    & 7.70 & 7.16  & 6.89 & 5.76 & 5.99 & 6.12 \\
      S$_3$    & 8.38 & 8.02  & 7.66 & 6.34 & 6.53 & 6.68 \\
      S$_4$    & 8.60 & 8.19  & 8.00 & 6.50 & 6.64 & 6.81 \\
      S$_5$    & 8.63 & 8.37  & 8.19 & 6.51 & 6.68 & 6.84 \\ 
    \hline
     \multicolumn{7}{c}{TDDFT(PBE0)}\\
    \hline
    State &  cc-pVDZ & cc-pVTZ & cc-pVQZ & aug-cc-pVDZ & aug-cc-pVTZ & aug-cc-pVQZ \\ 
    \hline
      S$_1$    & 4.34  & 4.35 & 4.34 & 4.31 & 4.33 & 4.33  \\
      S$_2$    & 7.54  & 7.06 & 6.78 & 6.08 & 6.10 & 6.09  \\
      S$_3$    & 7.74  & 7.63 & 7.58 & 6.68 & 6.67 & 6.66  \\
      S$_4$    & 7.86  & 7.78 & 7.62 & 6.79 & 6.77 & 6.75  \\
      S$_5$    & 8.20  & 8.00 & 7.73 & 6.80 & 6.78 & 6.76  \\ 
    \hline
     \multicolumn{7}{c}{NEVPT2(6,6)}\\
    \hline
    State &  cc-pVDZ & cc-pVTZ & cc-pVQZ & aug-cc-pVDZ & aug-cc-pVTZ & aug-cc-pVQZ \\ 
    \hline
      S$_1$    &  4.47 & 4.50  & 4.52  & 4.55  & 4.55  & --  \\ 
      S$_2$    &  8.14 & 7.62  & 7.38  & 6.47  & 6.72  & --  \\ 
    \hline
    \end{tabular}
    \caption{Low lying singlet excited state of cyclobutantone calculated using ADC(2), LR-TDDFT(PBE0) and NEVPT2(6,6) as a function of basis set at the ground state optimised geometry. For the NEVPT2, due to the size of the active space used only the S$_1$ and S$_2$ are included in the calculations.} 
    \label{tab:es}
\end{table}

\begin{table}[h]
    \centering
    \begin{tabular}{|c|c|c|c|c|c|c|c|}
    \hline
    State & NEVPT2(2,2) & NEVPT2(4,4) & NEVPT2(6,6) & NEVPT2(8,8) & NEVPT2(12,12) \\ 
    \hline
      S$_1$     &  4.42 & 4.36 & 4.55 & 4.65 &  4.48    \\
      S$_2$     & 10.68 & 8.57 & 6.47 & 6.51 &  6.43   \\
    \hline
    \end{tabular}
    \caption{Low lying singlet excited state of cyclobutantone calculated using NEVPT2 as a function of size of the active space calculated using the aug-cc-pVDZ basis set.}
    \label{tab:es}
\end{table}

\clearpage
\newpage
\subsection{ADC(2) and LR-TDDFT Excited State Energies at S$_1$ and S$_2$ Optimised Geometries}

\begin{table}[h]
    \centering
    \begin{tabular}{|c||c|c||c|c||}
         \hline
         State & \multicolumn{2}{|c||}{ADC(2)} & \multicolumn{2}{|c||}{LR-TDDFT(PBE0)}  \\
   \hline
    & Energy / eV & Character (Irrep.) & Energy / eV & Character (Irrep.)  \\
   \hline
        GS      &  1.13 &  ---   (A')  & 0.93 & ---   (A')  \\   
        S$_1$   &  3.96 &  H$\rightarrow$L   (A")  & 4.02  & H$\rightarrow$L   (A")  \\
        S$_2$   &  6.96 &  H$\rightarrow$L+1 (A")  & 7.03  & H$\rightarrow$L+1 (A") \\
        S$_3$   &  7.36 &  H-2$\rightarrow$L (A")  & 7.17  & H-2$\rightarrow$L (A") \\
        S$_4$   &  7.49 &  H-1$\rightarrow$L (A')  & 7.30  & H-1$\rightarrow$L (A')  \\
                &       &  H-4$\rightarrow$L       &       & H-4$\rightarrow$L    \\
        T$_1$   &  3.56 &  H$\rightarrow$L   (A")  &  3.38 & H$\rightarrow$L   (A")\\
        T$_2$   &  5.99 &  H-1$\rightarrow$L (A')  &  5.71 & H-1$\rightarrow$L (A') \\
                &       &  H-4$\rightarrow$L       &       & H-4$\rightarrow$L \\
        T$_3$   &  6.93 &  H-2$\rightarrow$L (A")  &  6.78 & H-2$\rightarrow$L (A") \\
        T$_4$   &  7.30 &  H-3$\rightarrow$L (A')  &  7.19 & H-3$\rightarrow$L (A')\\
 \hline
    \end{tabular}
    \caption{Summary of electronic energies, characters, and symmetries of the ground- (GS) and electronically-excited (S$_{n}$/T$_{n}$; $n = [1..5]$) states of cyclobutanone evaluated at the S$_{1}$ minimum-energy geometry and at the ADC(2)/aug-cc-pVTZ and LR-TDDFT(PBE0)/aug-cc-pVDZ levels of theory. Comparative tables for evaluations at the S$_1$- (Table S3) and S$_2$-state (Table S4) minimum-energy geometries are presented in the SI. The relevant molecular orbitals are shown in the SI. The HOMO is designated as H; the LUMO is designated as L.}
    \label{tab:es}
\end{table}

\begin{table}[h]
    \centering
    \begin{tabular}{|c||c|c||c|c||}
         \hline
         State & \multicolumn{2}{|c||}{ADC(2)} & \multicolumn{2}{|c||}{LR-TDDFT(PBE0)}  \\
   \hline
    & Energy / eV & Character (Irrep.) & Energy / eV & Character (Irrep.)  \\
   \hline
        GS      &  0.51 &  ---   (A')              &  0.51 & ---   (A')  \\   
        S$_1$   &  4.79 &  H$\rightarrow$L   (A$_2$ [A"])  &  4.72  & H$\rightarrow$L   (A")  \\
        S$_2$   &  5.86 &  H$\rightarrow$L+1 (B$_1$ [A"])  &  5.84   & H$\rightarrow$L+1 (A") \\
        S$_3$   &  6.52 &  H$\rightarrow$L+4 (A$_2$ [A"])  &  6.54 & H$\rightarrow$L+3 (A") \\
        S$_4$   &  6.58 &  H$\rightarrow$L+2 (B$_1$ [A"])  &  6.56 & H$\rightarrow$L+2 (A') \\
        T$_1$   &  4.44 &  H$\rightarrow$L   (A$_1$ [A'])  &  4.13 & H$\rightarrow$L   (A")\\
        T$_2$   &  5.79 &  H-1$\rightarrow$L (A$_2$ [A"])  &  5.73 & H-1$\rightarrow$L (A') \\
                &       &  H-4$\rightarrow$L       &   & H-4$\rightarrow$L \\
        T$_3$   &  6.50 &  H$\rightarrow$L+1 (A$_2$ [A"])  &  6.44 &H$\rightarrow$L+1   (A")  \\
        T$_4$   &  6.54 &  H$\rightarrow$L+3 (A$_1$ [A'])  &  6.59 & H$\rightarrow$L+3 (A")\\
 \hline
    \end{tabular}
    \caption{ummary of electronic energies, characters, and symmetries of the ground- (GS) and electronically-excited (S$_{n}$/T$_{n}$; $n = [1..5]$) states of cyclobutanone evaluated at the S$_{2}$ minimum-energy geometry and at the ADC(2)/aug-cc-pVTZ and LR-TDDFT(PBE0)/aug-cc-pVDZ levels of theory. Comparative tables for evaluations at the S$_1$- (Table S3) and S$_2$-state (Table S4) minimum-energy geometries are presented in the SI. The relevant molecular orbitals are shown in the SI. The HOMO is designated as H; the LUMO is designated as L.}
    \label{tab:es}
\end{table}

\clearpage
\newpage

\subsection{Molecular Orbitals of Cyclobutanone}

\begin{figure}[h]
    \centering
    \begin{tabular}{cccc}
    \includegraphics[width=0.2\linewidth]{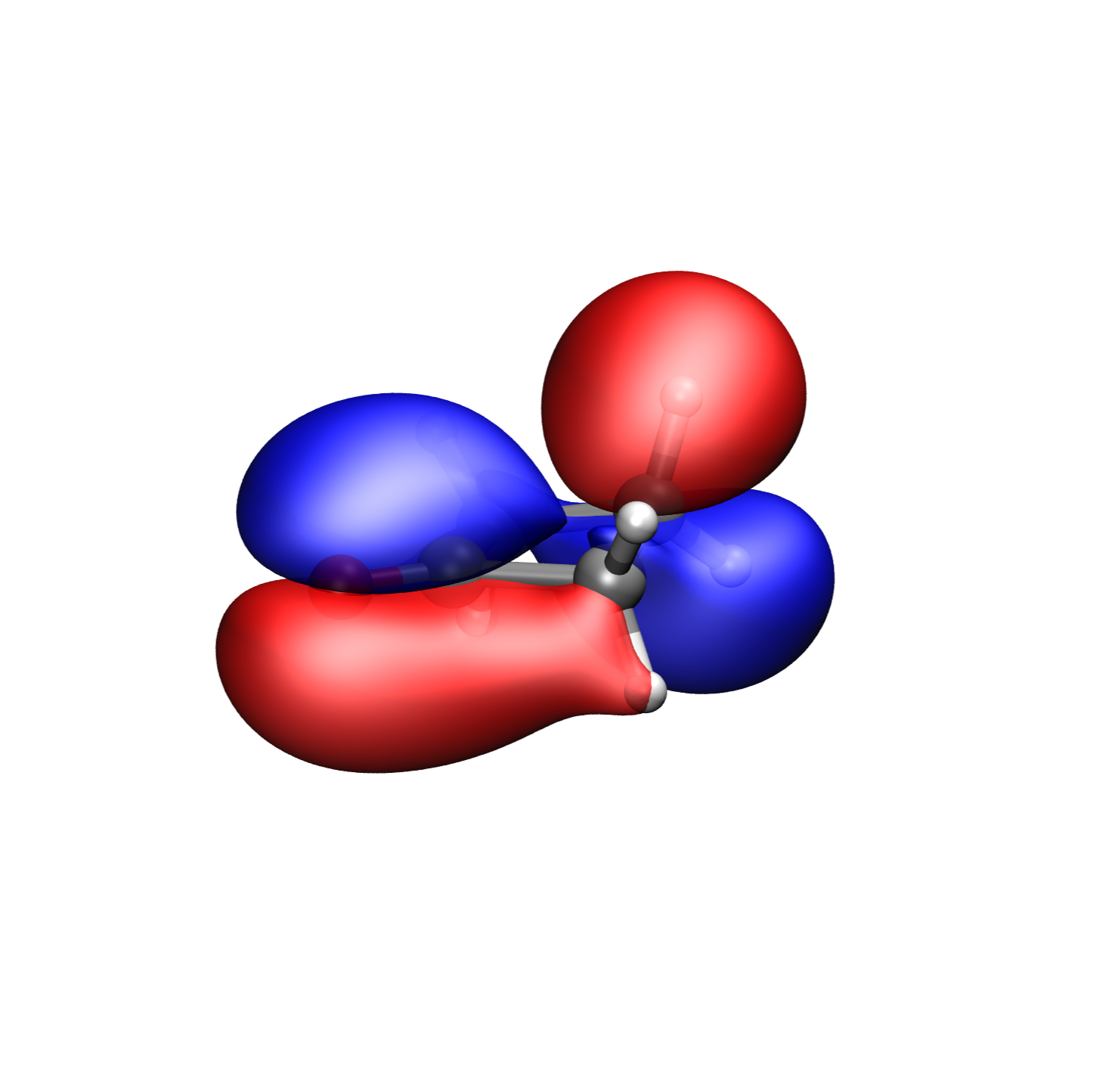} & \includegraphics[width=0.2\linewidth]{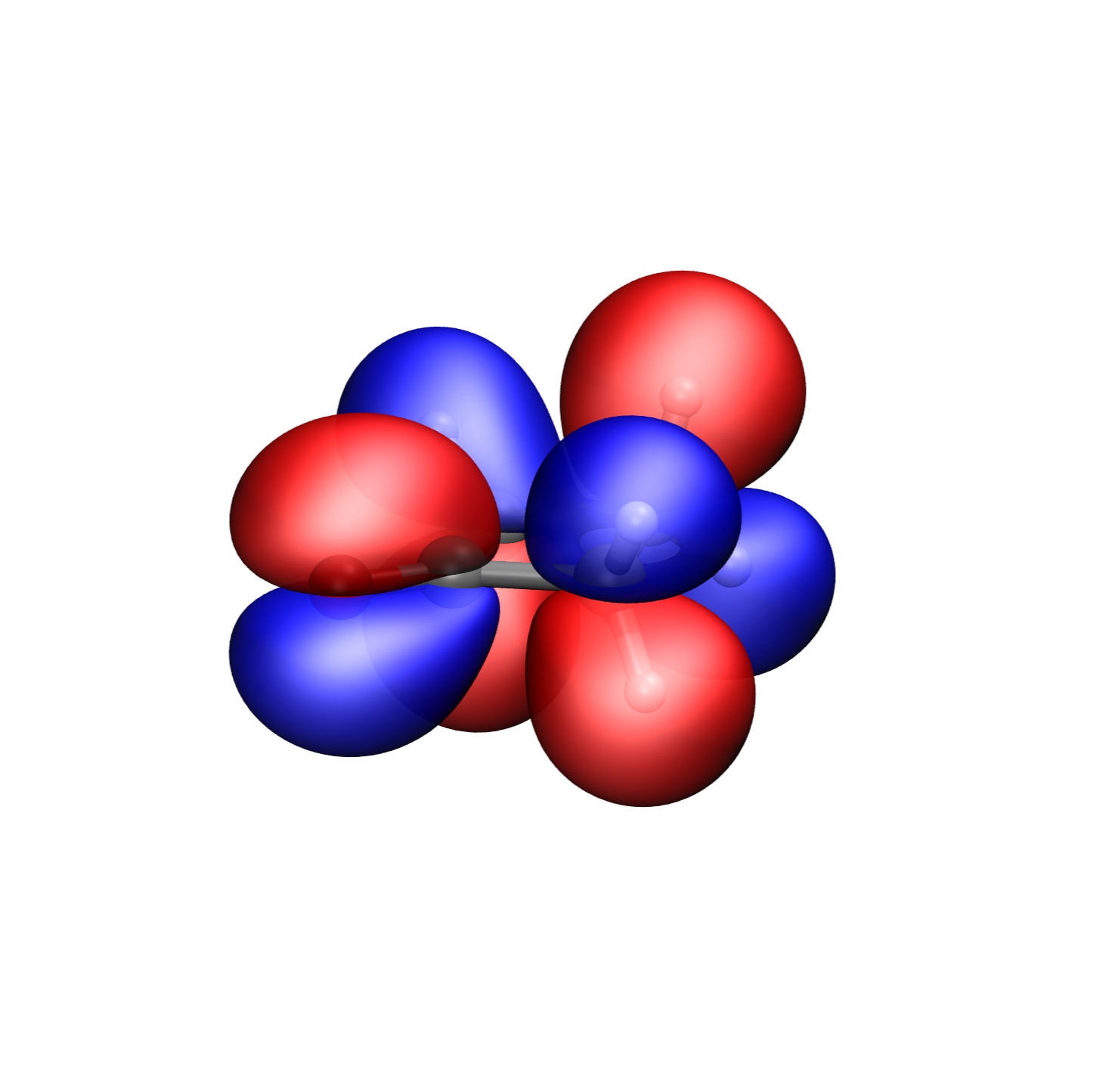} & \includegraphics[width=0.2\linewidth]{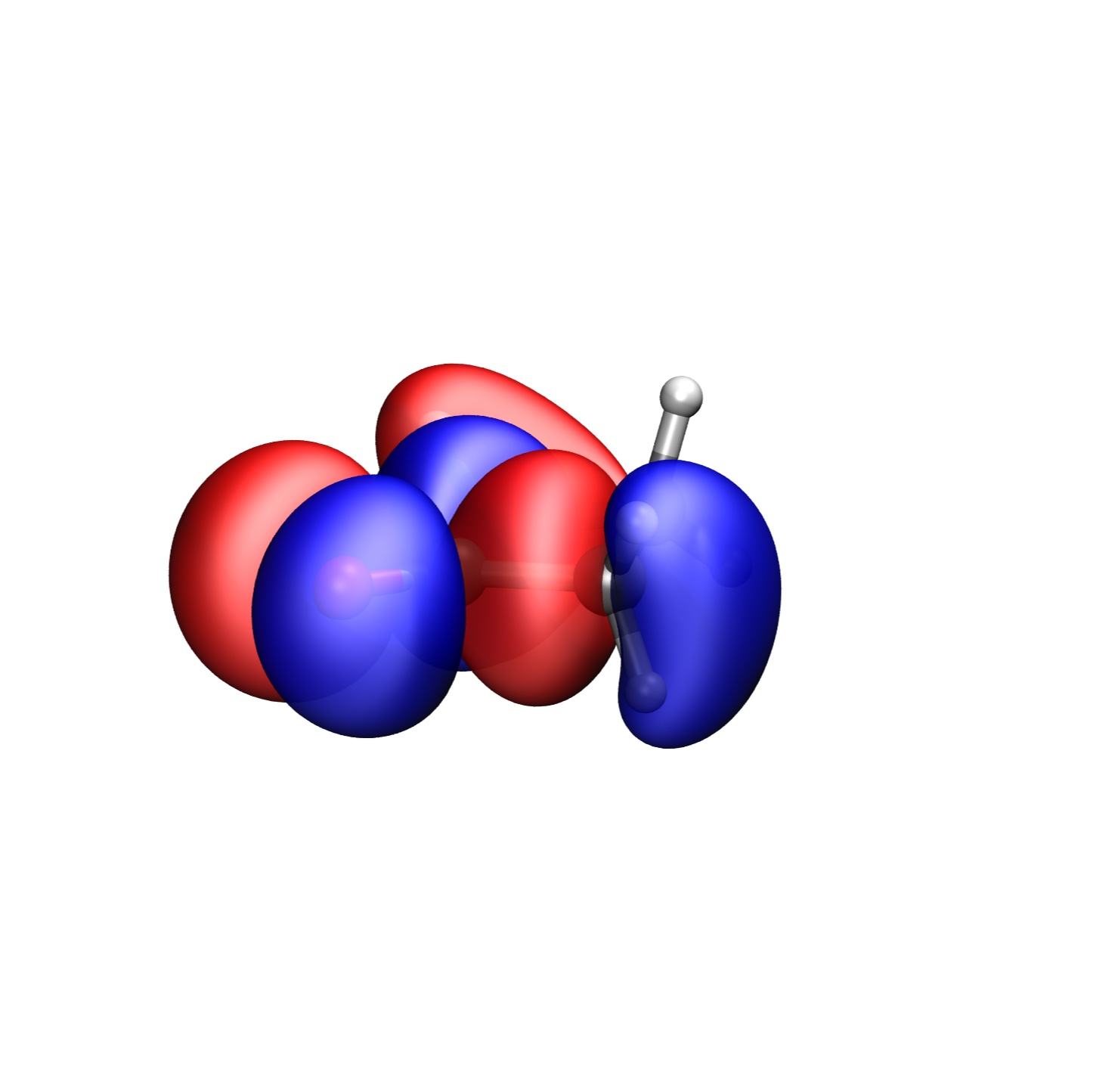} & \includegraphics[width=0.2\linewidth]{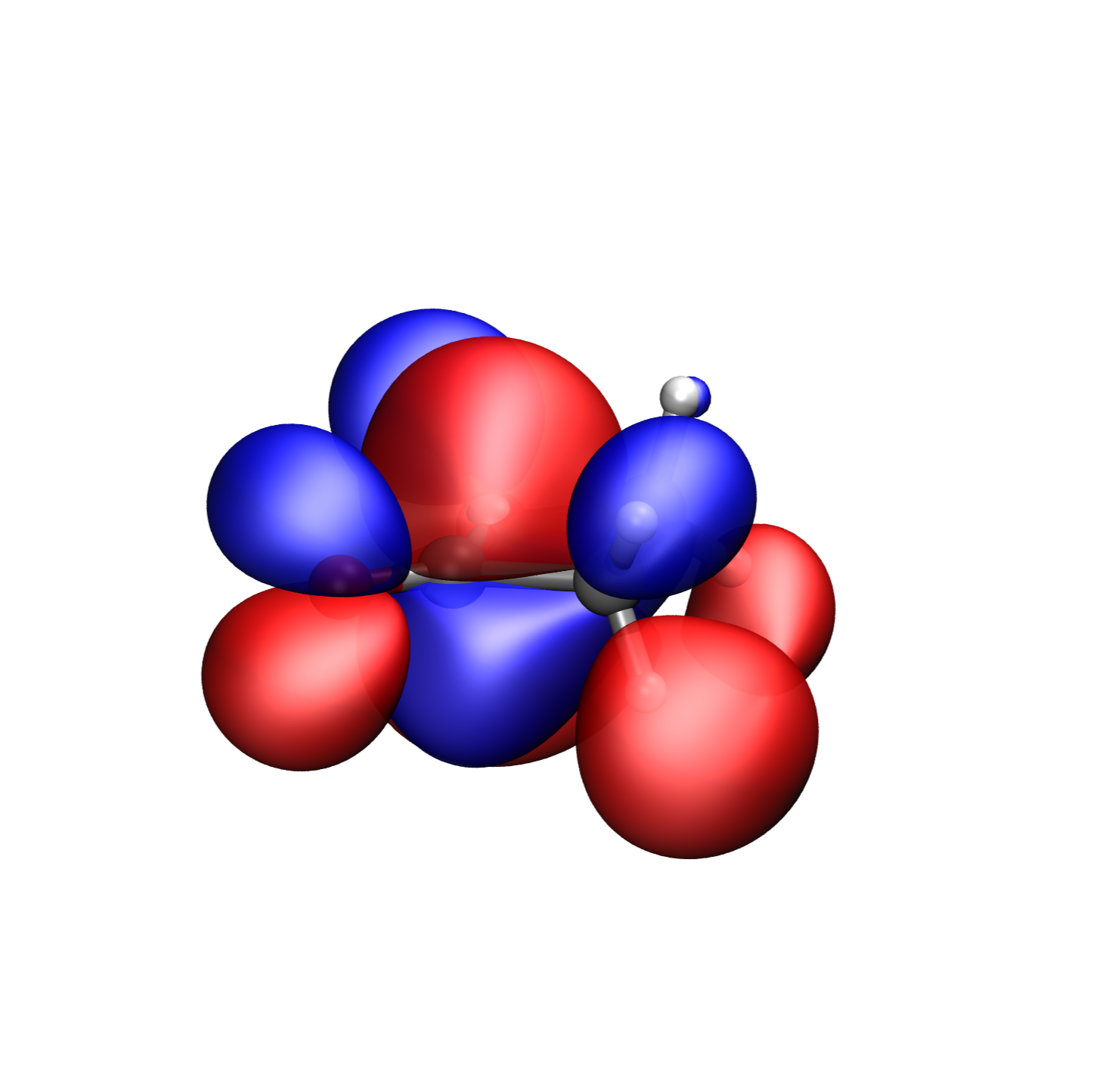} \\ 
    HOMO-4 & HOMO-1 & HOMO & LUMO\\
    \includegraphics[width=0.2\linewidth]{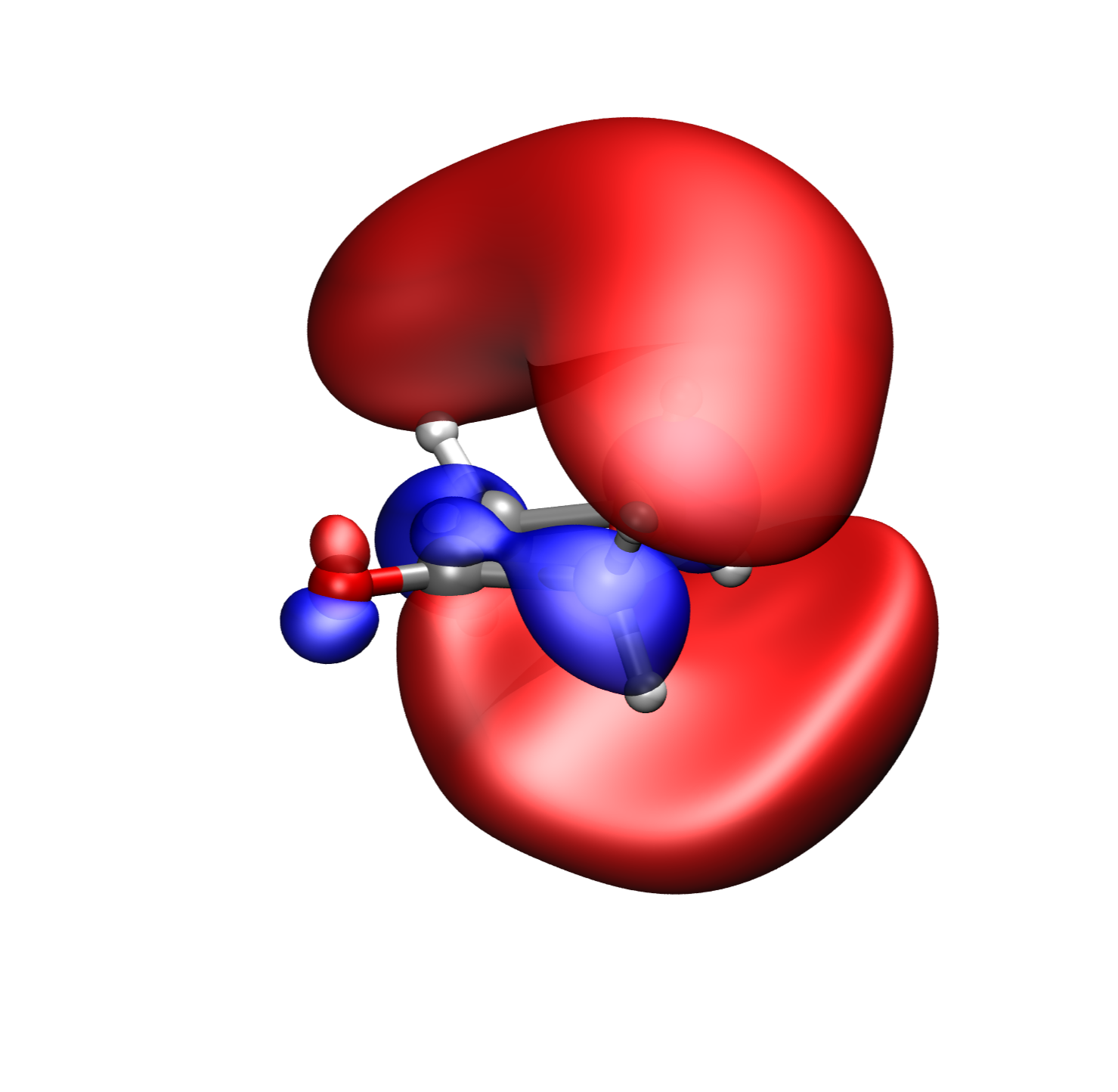} &
    \includegraphics[width=0.2\linewidth]{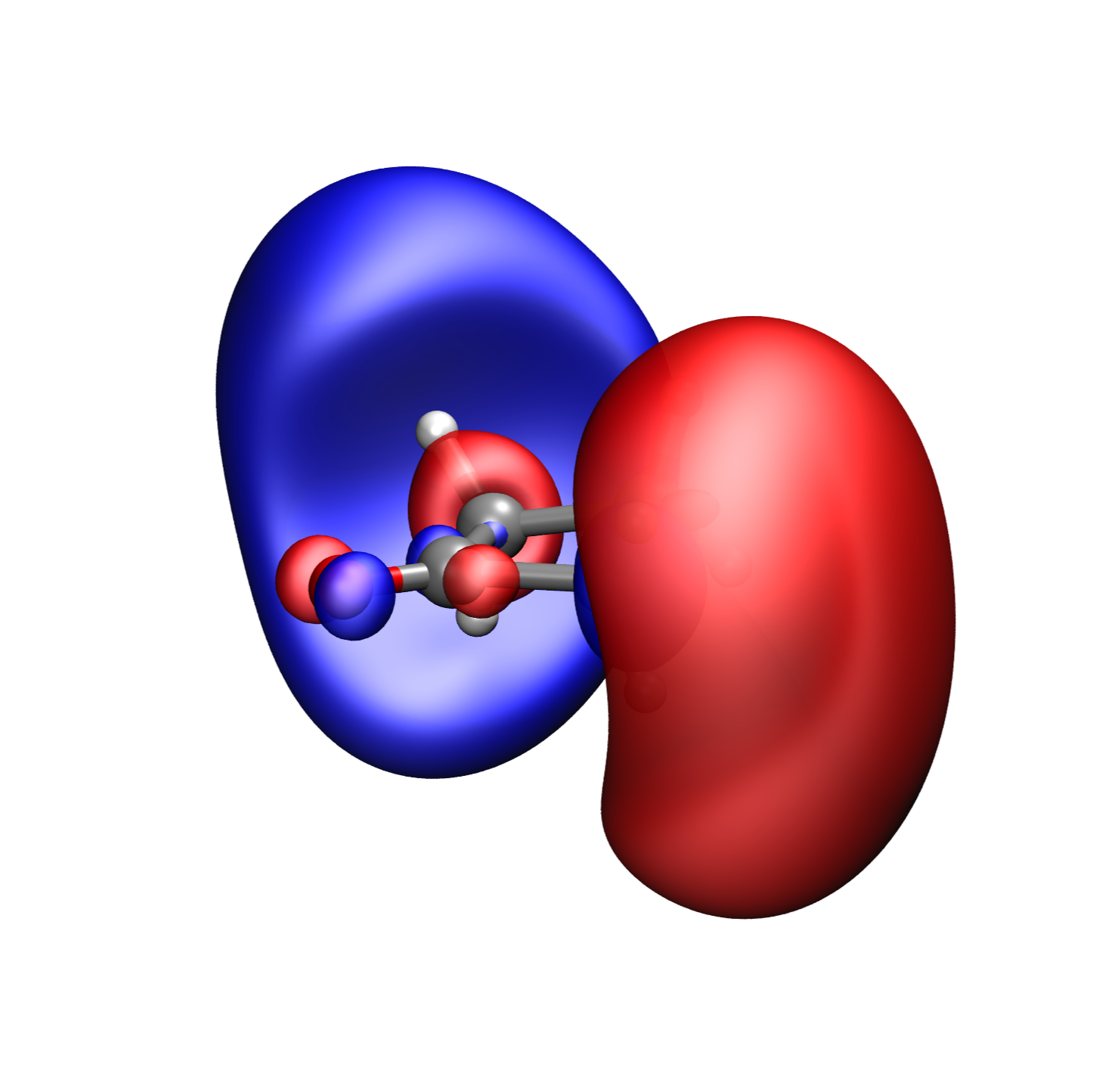} &
    \includegraphics[width=0.2\linewidth]{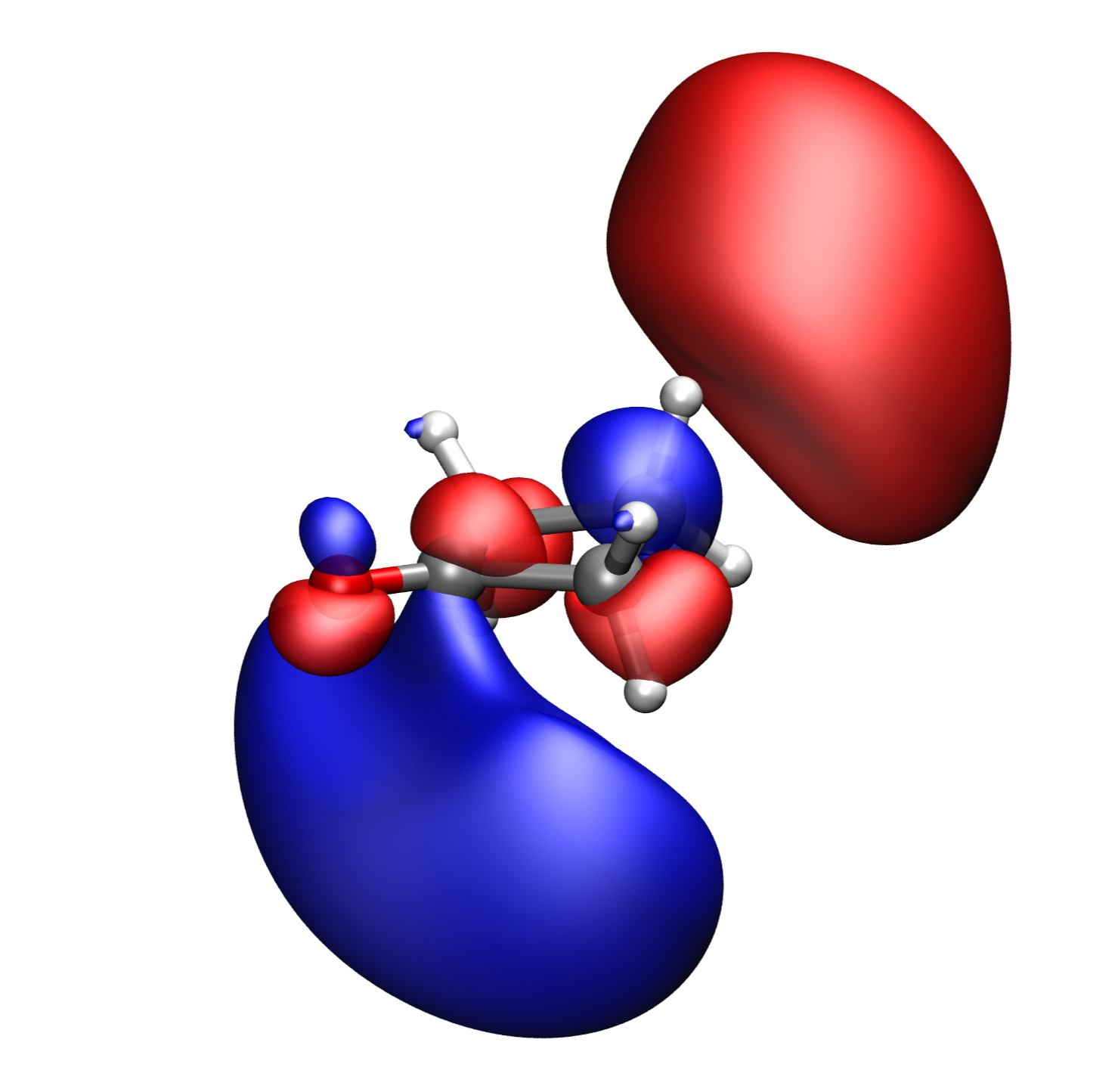}  &
    \includegraphics[width=0.2\linewidth]{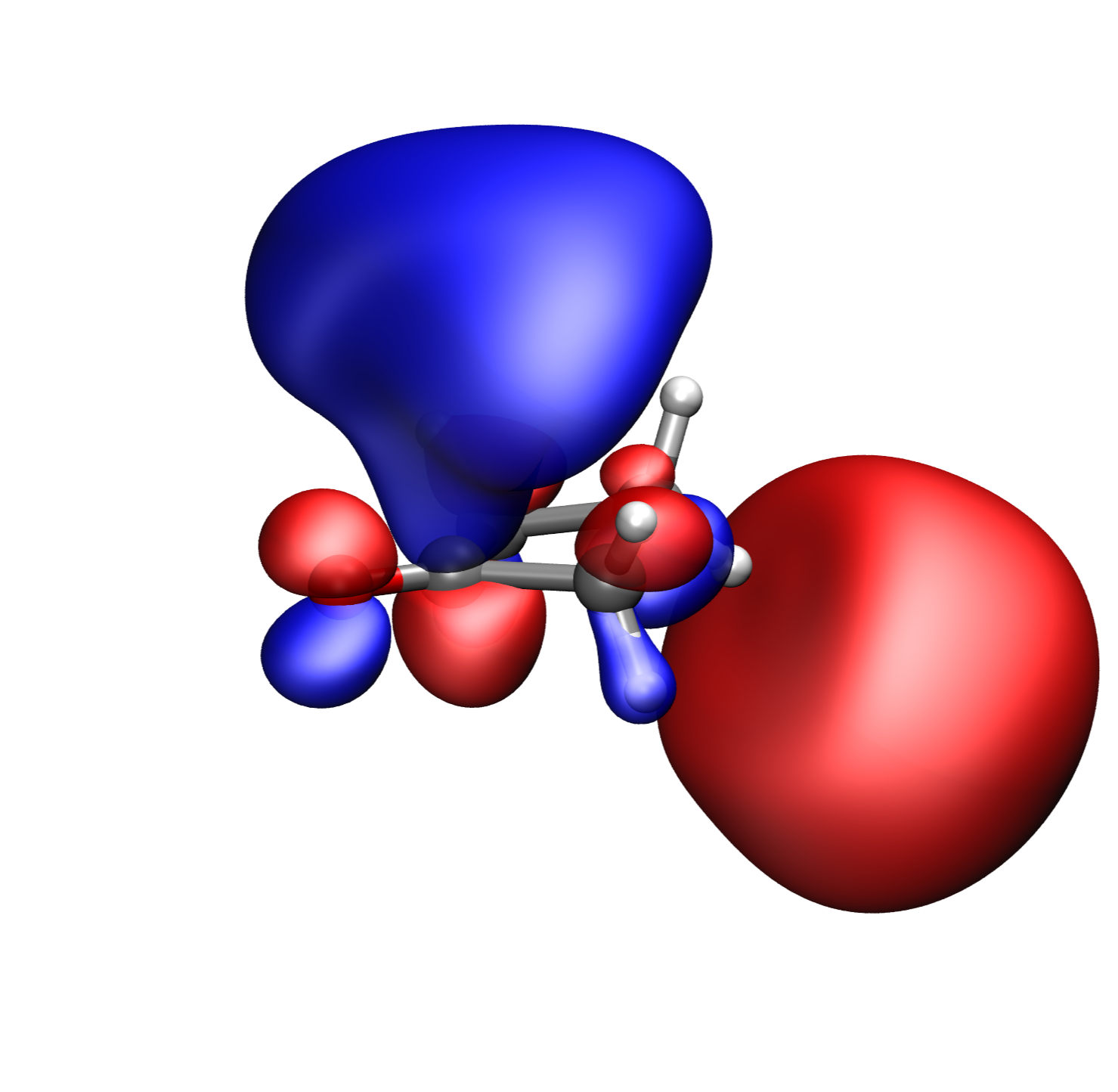}  \\
     LUMO+1 & LUMO+2 & LUMO+3 & LUMO+4\\
    \end{tabular}
    \caption{Molecular orbitals of cyclobutanone involved in the lowest excited states reported here and in the main text.}
    \label{fig:mos}
\end{figure}

\clearpage
\newpage

\subsection{Potential Energy Curve along C=O Breathing Mode}

\begin{figure}[h]
\centering
\includegraphics[width=0.7\linewidth]{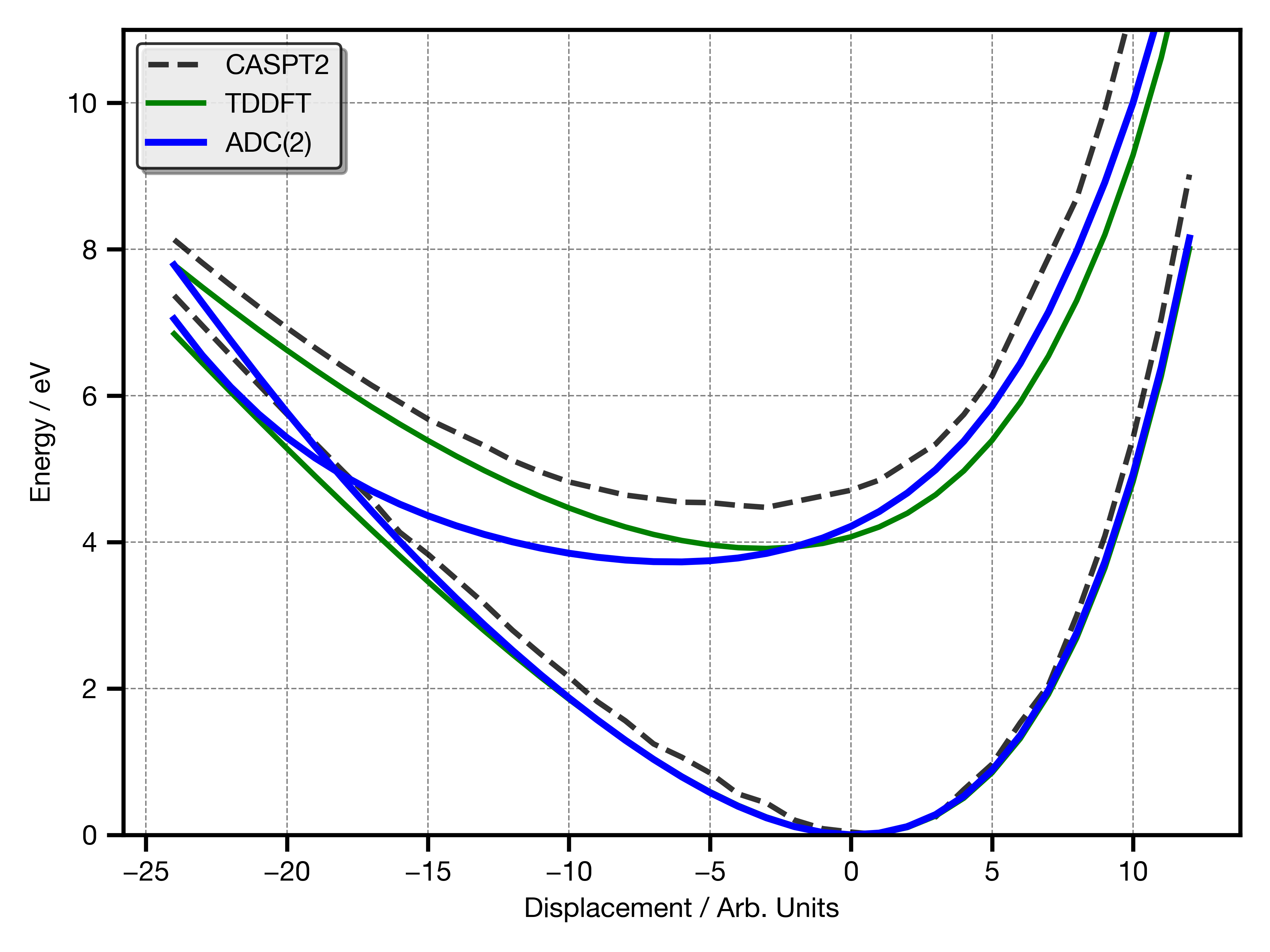}
\caption{S$_0$/S$_1$ potential Energy Surface along $\nu_{21}$, C=O stretching normal mode using NEVPT2(12,12)/aug-cc-pvtz (black dashed), LR-TDDFT(PBE0)/aug-cc-pvtz (green) and ADC(2)/aug-cc-pvtz (blue dashed) levels of theory. This clearly shows the artificial crossing for ADC(2) consistent with the observations of ref. \cite{marsili2021caveat}.}
\label{}
\end{figure}

\clearpage
\newpage
\subsection{Linear Reaction Coordinates leading to Conical Intersections}

\begin{figure}[h]
\centering
\includegraphics[width=0.6\linewidth]{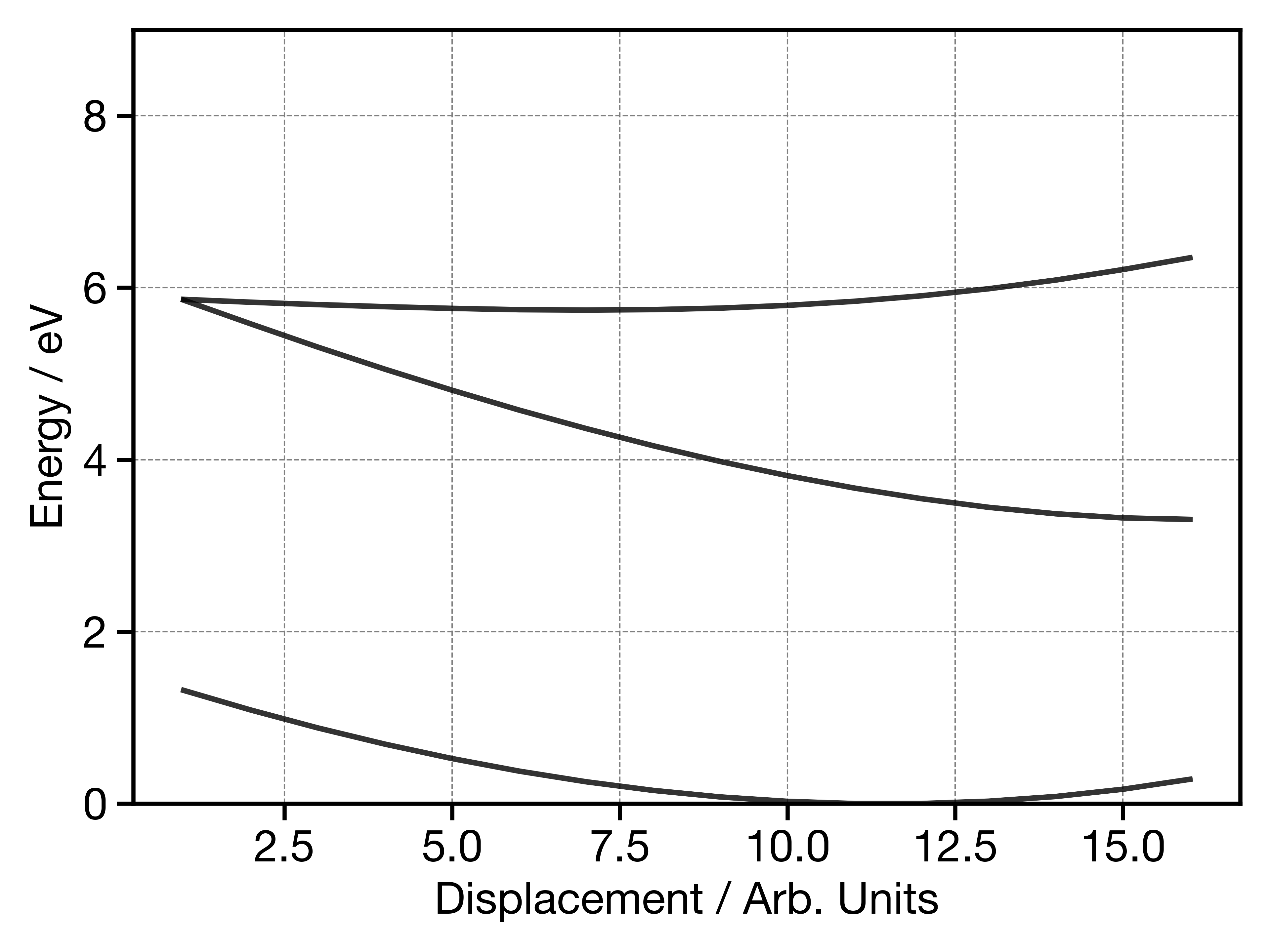}
\caption{DFT/LR-TDDFT(PBE0)/aug-cc-pvdz potential energy surface along a linear reaction coordinate from the Franck-Condon to the conical intersection between the S$_2$/S$_1$ states.}
\label{}
\end{figure}

\begin{figure}[h]
\centering
\includegraphics[width=1.0\linewidth]{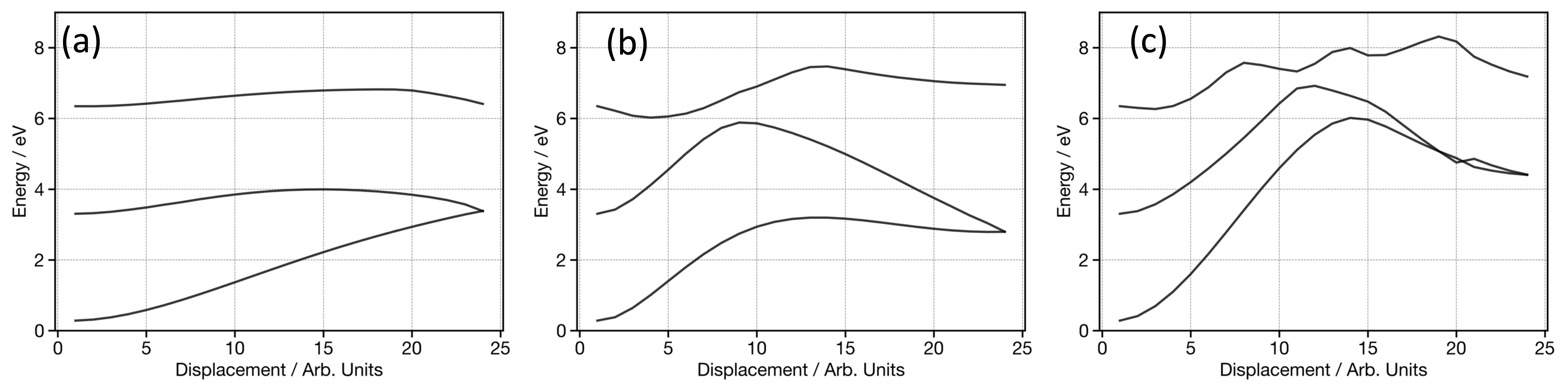}
\caption{DFT/LR-TDDFT(PBE0)/aug-cc-pvdz potential energy surface along a linear reaction coordinate from (a) the S$_1$ minimum energy geometry to C$_{\alpha}$ conical intersection between the S$_1$/S$_0$ states, (b) the S$_1$ minimum energy geometry to C$_{\beta}$ conical intersection between the S$_1$/S$_0$ states and (c) the S$_1$ minimum energy geometry to C$_{\alpha}$, C$_{\beta}$ conical intersection between the S$_1$/S$_0$ states. Structures are shown in the main text and the cartesian coordinates are provided below.}
\label{}
\end{figure}

\clearpage
\newpage
\section{Quantum Dynamics}
\subsection{Normal Modes}

\begin{figure}[h]
    \centering
    \begin{tabular}{cccc}
    \includegraphics[width=0.23\linewidth]{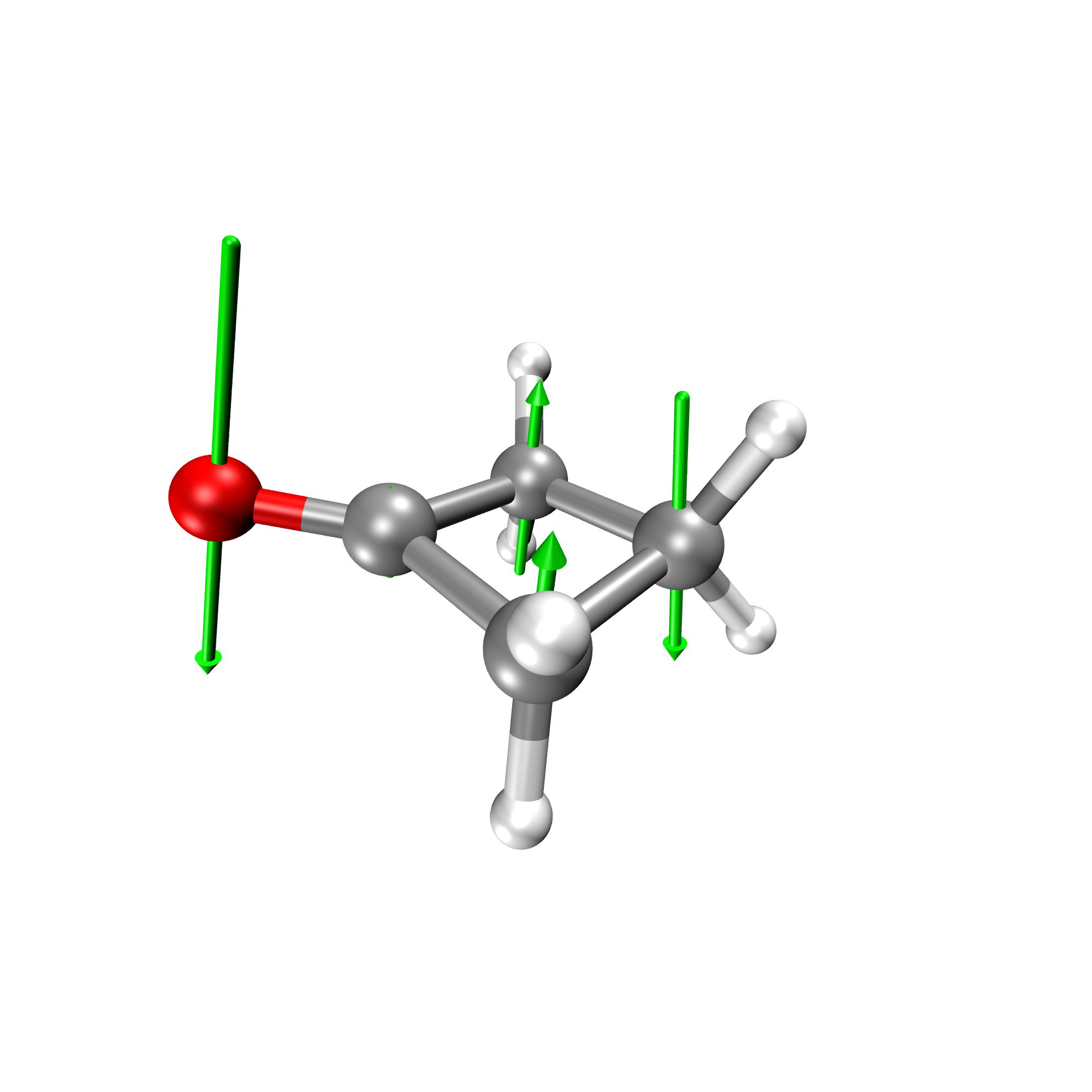} & \includegraphics[width=0.23\linewidth]{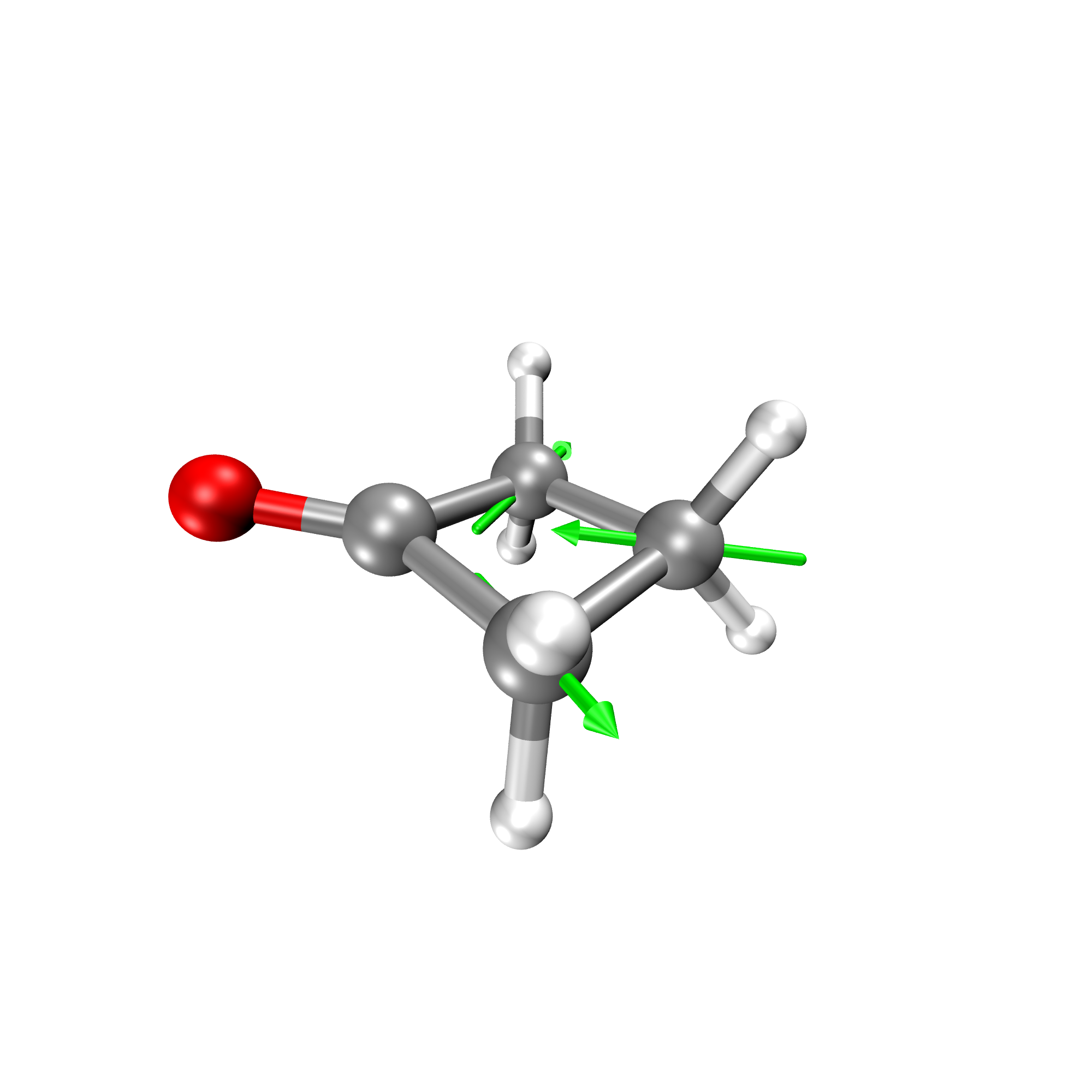} & \includegraphics[width=0.23\linewidth]{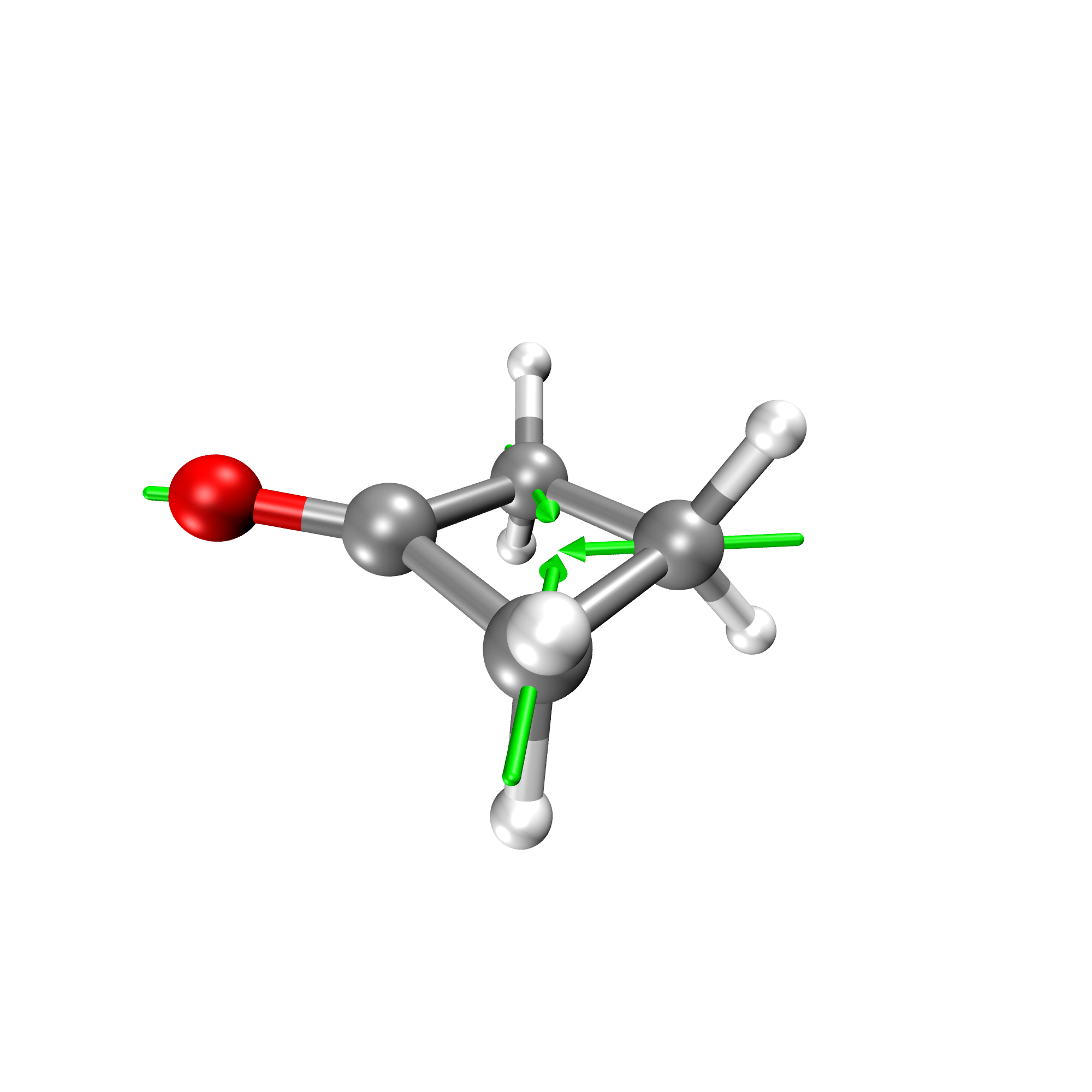} & \includegraphics[width=0.23\linewidth]{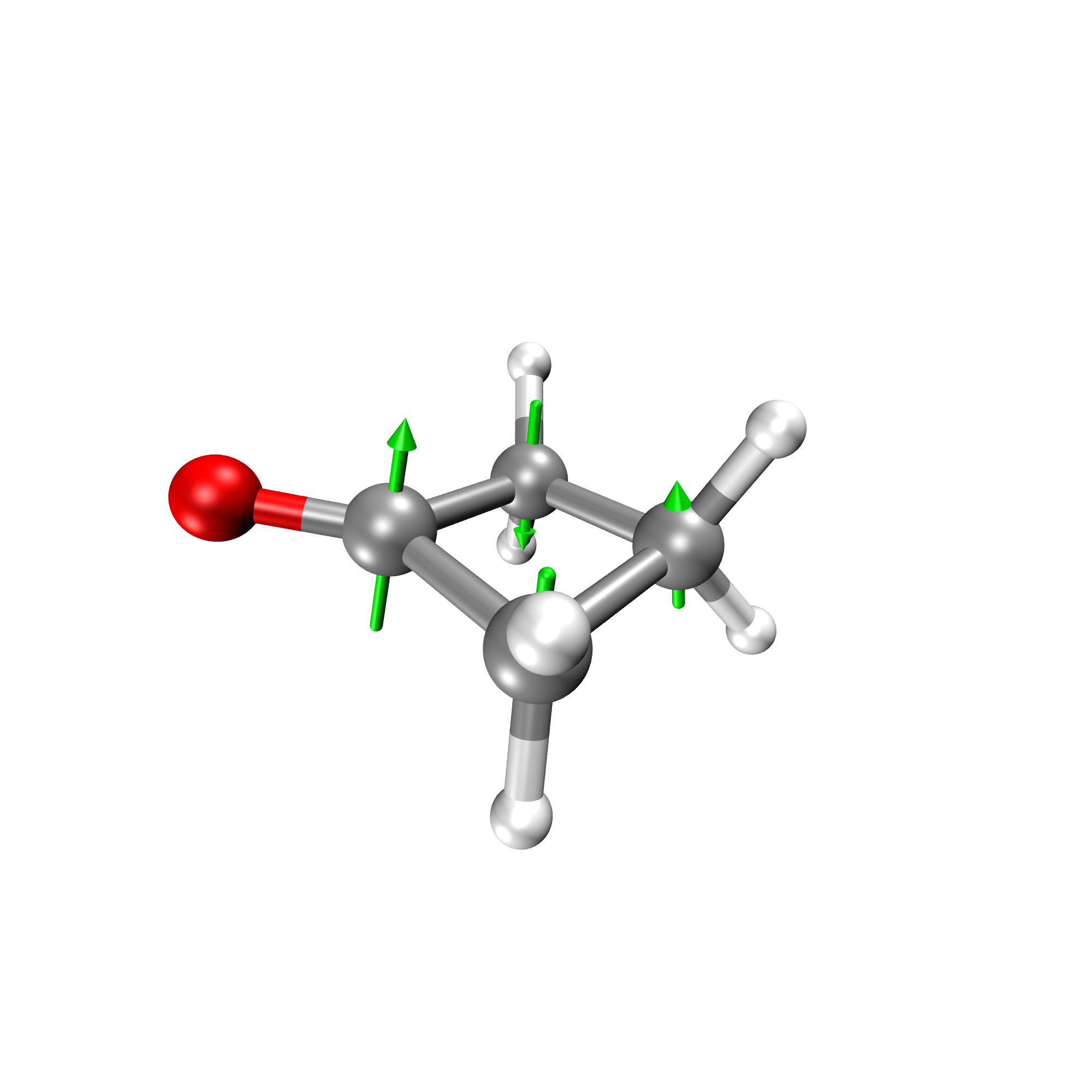} \\ 
    $\nu_{1}$ & $\nu_{7}$ & $\nu_{10}$ & $\nu_{11}$\\
    \includegraphics[width=0.23\linewidth]{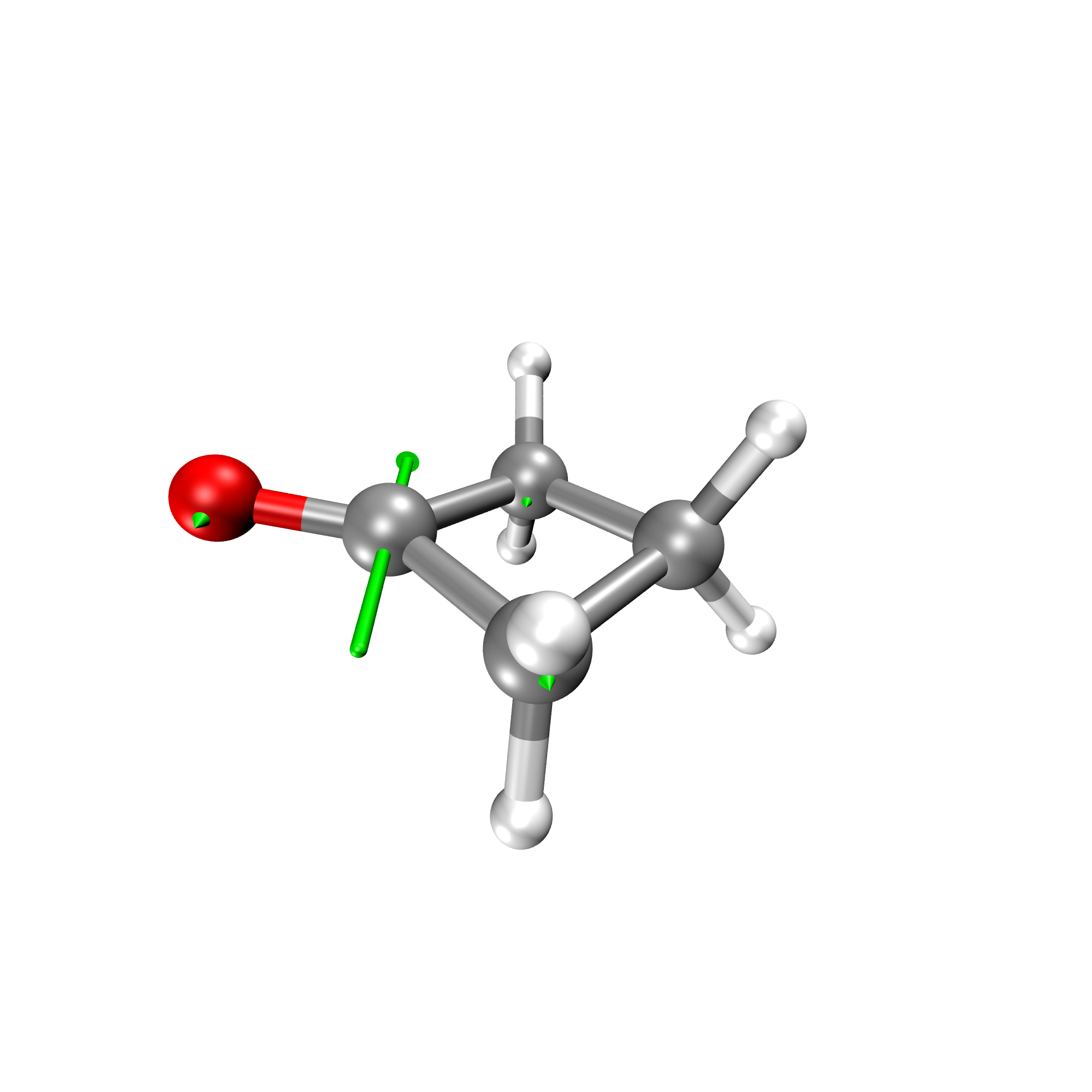} &
    \includegraphics[width=0.23\linewidth]{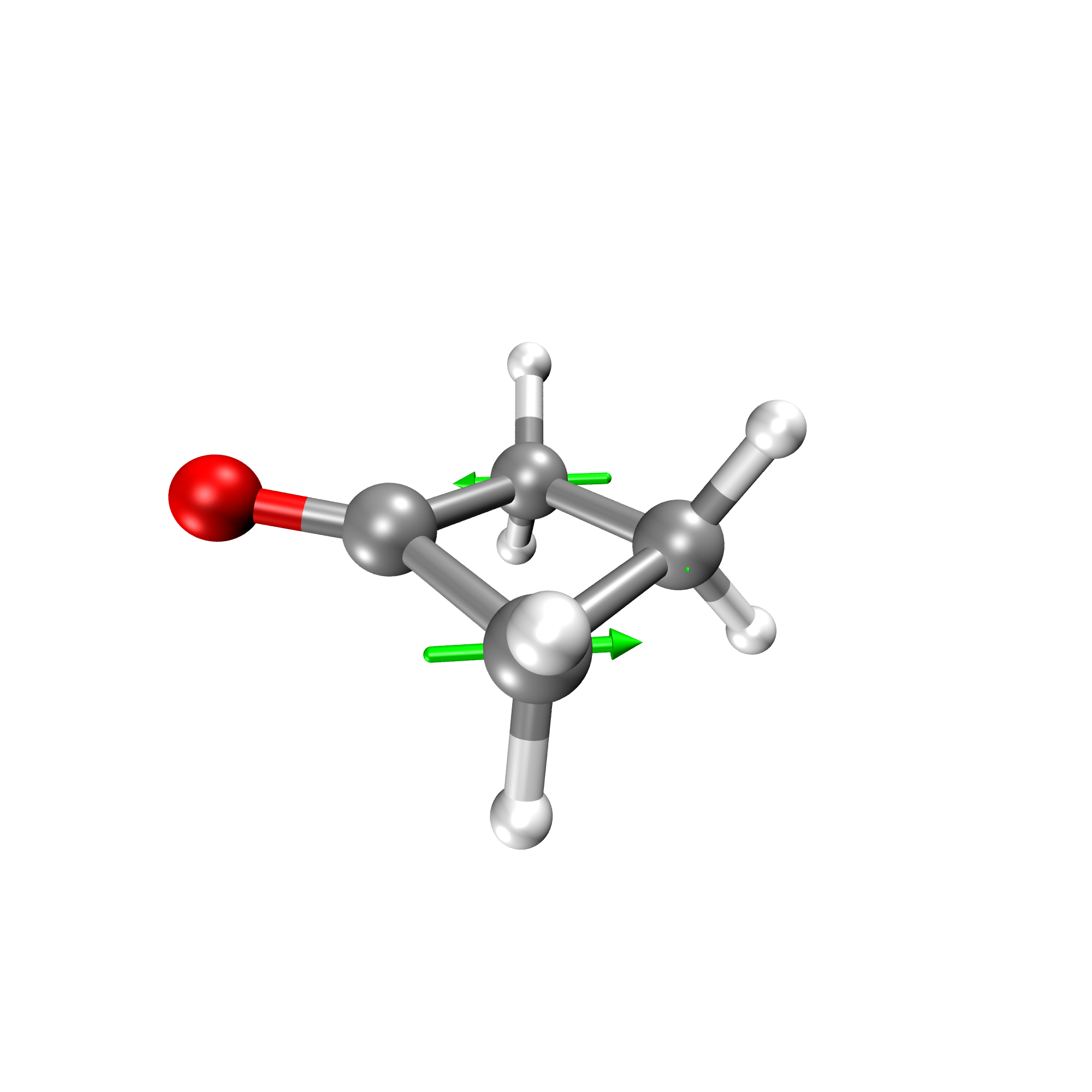} &
    \includegraphics[width=0.23\linewidth]{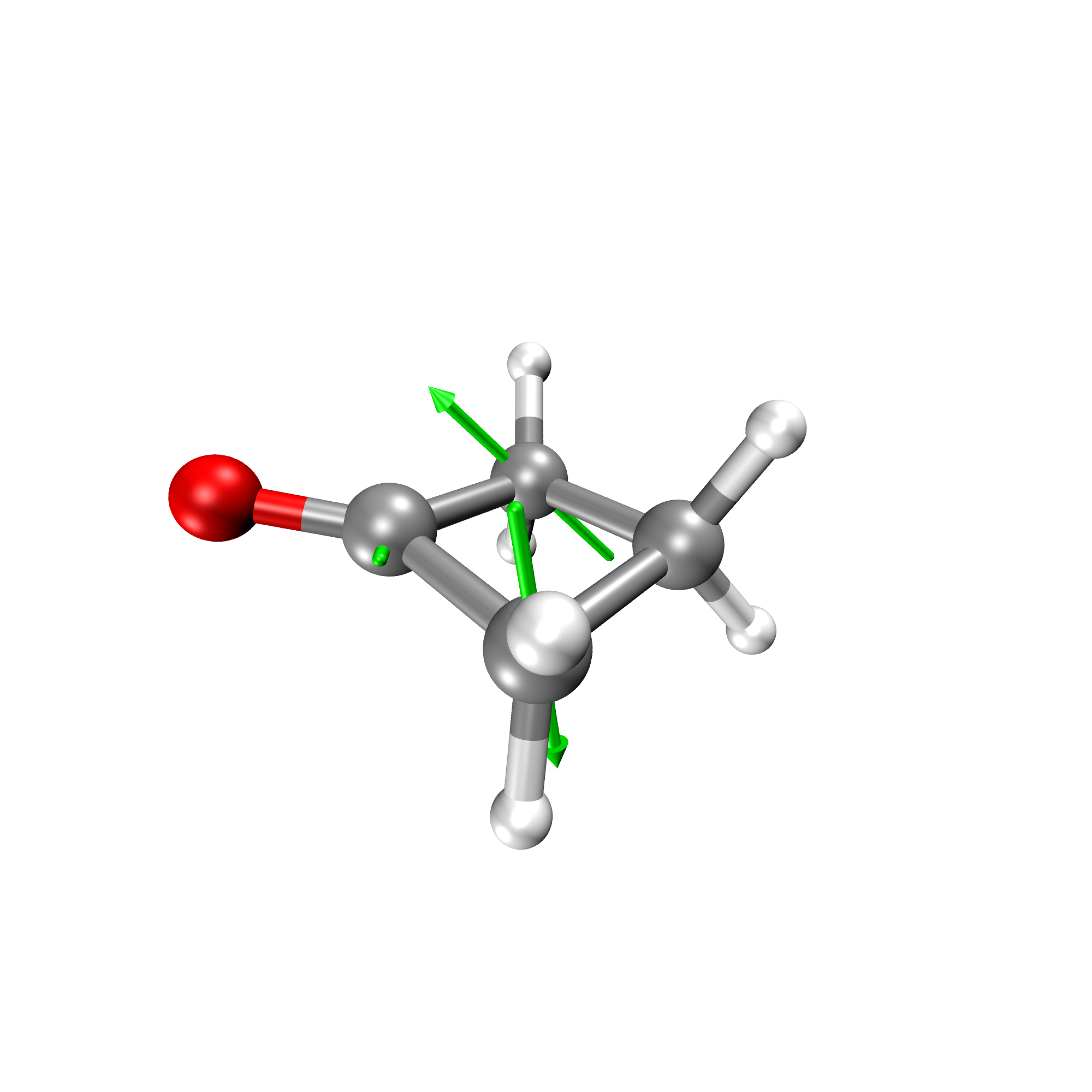}  &
    \includegraphics[width=0.23\linewidth]{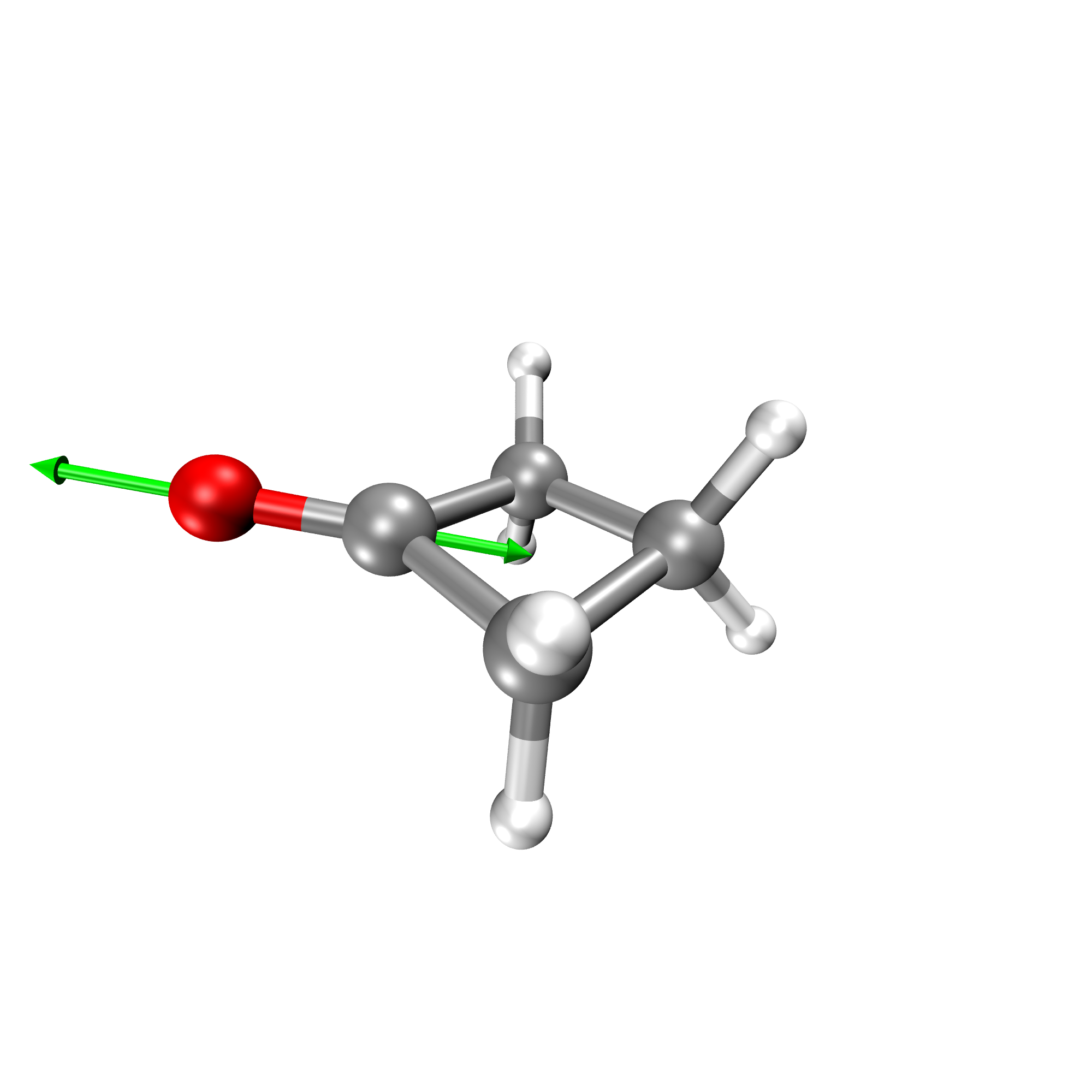}  \\
     $\nu_{12}$ & $\nu_{13}$ & $\nu_{15}$ & $\nu_{21}$\\
    \end{tabular}
    \caption{Normal modes of cyclobutanone included in the model Hamiltonian used within this work.}
    \label{fig:mos}
\end{figure}

\clearpage
\newpage

\subsection{Model Hamiltonian Expansion Coefficients}

\begin{table}[h]
    \centering
    \begin{tabular}{|c|ccccc|ccccc|}
    \hline
        \multirow{2}{*}{Mode (a')}    & \multicolumn{10}{c|}{$\kappa$ / eV} \\
        \cline{2-11}
         & \multicolumn{5}{c|}{Singlets} &  \multicolumn{5}{c|}{Triplets} \\   \hline
         & GS & $S_1$ & $S_2$ & $S_3$& $S_4$ & $T_1$ & $T_2$ & $T_3$ & $t_4$ & $T_5$ \\
        $Q_1$   & 0.000  &  0.052   &  0.024     &  0.011     &  0.024     &  0.055     &  0.016     &  0.055     &-0.017       & 0.023  \\
        $Q_7$   & 0.000  &  -0.001   &  0.022     &  0.016     &  0.010     &   0.018    &  0.036     &   0.029    &  0.031     &  -0.003 \\
        $Q_{12}$ & 0.000  &  0.050   &   -0.028    &  -0.017     &  -0.001     &  0.020     &  0.030     & 0.030      &  0.030     &  0.040 \\
        $Q_{15}$ & 0.000  &  -0.006   & -0.128      &  -0.136     &  -0.116     &  0.002     &  -0.128     & 0.049      &  -0.131     & -0.114  \\
        \hline
                \multirow{2}{*}{Mode (a')}    & \multicolumn{10}{c|}{$\gamma$ / eV} \\
        \cline{2-11}
         & \multicolumn{5}{c|}{Singlets} &  \multicolumn{5}{c|}{Triplets} \\  \hline 
         & GS & $S_1$ & $S_2$ & $S_3$& $S_4$ & $T_1$ & $T_2$ & $T_3$ & $t_4$ & $T_5$ \\
        $Q_1$ & 0.010  &  0.001   & 0.007   &  0.005     &   0.007    &  0.005     &  0.009     &  -0.011     &  0.005     &  0.004 \\
        $Q_7$ & -0.026  &  -0.027   &  -0.028     &  -0.029     &  -0.029     &  -0.027     &  -0.027     &  -0.028     &  -0.029     &  -0.030 \\
        $Q_{12}$ & 0.001  &  -0.022   &  -0.009     &  -0.009     &  -0.009     &  -0.020     &  -0.020     &  -0.020     &  -0.020     & 0.020  \\
        $Q_{15}$ & 0.069 &  -0.074   &  -0.073 & -0.073     &  -0.070     &  -0.073     &  -0.077     &  -0.092     &   -0.078    &  -0.074 \\
        \hline
                \multirow{2}{*}{Mode (a')}    & \multicolumn{10}{c|}{$\delta$ / eV} \\
        \cline{2-11}
         & \multicolumn{5}{c|}{Singlets} &  \multicolumn{5}{c|}{Triplets} \\  \hline
         & GS & $S_1$ & $S_2$ & $S_3$& $S_4$ & $T_1$ & $T_2$ & $T_3$ & $t_4$ & $T_5$ \\
        $Q_1$ & 0.002  &  0.002   &   0.002    & 0.002      &  0.002     &  0.002      &  0.002     &  0.003     &  0.002     &  0.003 \\
        $Q_7$ & ---  &  ---   &  ---     &  ---     &   ---    &  ---     &  ---     &    ---   &    ---   & ---  \\
        $Q_{12}$ &  --- &  ---   &   ---    &   ---    &       ---&   ---    &  ---     &  ---     &      --- &  --- \\
        $Q_{15}$ & ---  &  ---   &  ---     &  ---     &    ---   &  ---     &    ---   &---       &---       & ---  \\        
          \hline
    \end{tabular}
    \caption{Expansion Coefficients used for the model Hamiltonian used within this work.}
    \label{tab:QVC_a}
\end{table}

\begin{table}[h]
    \centering
    \begin{tabular}{|c|ccccc|ccccc|}
    \hline
         $Q_{10}$& \multicolumn{5}{c|}{Singlets} &  \multicolumn{5}{c|}{Triplets} \\   \hline
          & GS & $S_1$ & $S_2$ & $S_3$& $S_4$ & $T_1$ & $T_2$ & $T_3$ & $t_4$ & $T_5$ \\
        D               & 11.714  &  2.642   &   2.547    &  2.627     &   2.382    &  5.167     &  2.464     &  13.518     &  3.432     & 2.000   \\   
        $\alpha$        & 0.101  & 0.117    & 0.163      &   0.161    &  0.161     &   0.102    &  0.167     &   0.066    &   0.143    &  0.174 \\
        $x_e$           & 0.00  & 3.008    &  0.158     &   0.205    &  0.254     &  2.363     &  0.006     & 4.126      &  0.158     & 0.110  \\
        $\delta \varepsilon$    & 0.000  & -0.484    &   -0.035    &   -0.026    &   -0.020    &    -0.441   &  -0.027     &  1.351     &  -0.059     &  -0.002  \\
        \hline
         $Q_{21}$& \multicolumn{5}{c|}{Singlets} &  \multicolumn{5}{c|}{Triplets} \\   \hline
          & GS & $S_1$ & $S_2$ & $S_3$& $S_4$ & $T_1$ & $T_2$ & $T_3$ & $t_4$ & $T_5$ \\
        D               &  20.890 &  17.810   &  18.950     &   9.640    &  15.970     &  17.910     &  18.880     & 18.100      &   7.070    & 14.090  \\   
        $\alpha$        &  0.055 &  0.057   &  0.056     &   0.072    &   0.060    &  0.057     &  0.056     &   0.056    &   0.080    &  0.063 \\
        $x_e$           & 0.000  & -0.064    &  -0.189     &  -0.011     &  -0.084     &  -0.122     &  -0.186     &  0.084     &   0.080    &   -0.053\\
        $\delta \varepsilon$    &  0.000 &  0.008   & -0.009      &  -0.017     &   -0.017    &   0.003    &  0.005     &   0.002    &    -0.006   &  0.001 \\
        \hline
    \end{tabular}
    \caption{Expansion Coefficients used for the model Hamiltonian used within this work.}
    \label{tab:QVC2_a}
\end{table}

\begin{table}[h]
    \centering
    \begin{tabular}{|c|cccc|cccc|}
    \hline
       $\eta$ / cm$^{-1}$& T$_1$ &  T$_2$ &  T$_3$ &  T$_4$ &  S$_1$ &   S$_2$ &    S$_3$ &    S$_4$  \\
       \hline
         T$_1$          &  ---     &  54.49 &  0.32  &  2.19  &   0.02 &  0.43   &  0.94    &   0.00    \\
         T$_2$          & 54.49 & ---       &  0.74  &  3.30  &   0.00 &  0.00   &   0.00   &  -0.20    \\
         T$_3$          & 0.32  & 0.74   & ---       &  0.05  &   0.14 &  0.01   &  -0.04   &   0.00    \\
         T$_4$          & 2.19  & 3.30   &  0.05  &  ---      &   -1.48&  0.03   &   0.01   &   0.00    \\
         \hline
    \end{tabular}
    \caption{Spin-orbit coupling $\eta$ in cm$^{-1}$ between the electronic states included in the Hamiltonian.}
    \label{tab:soc}
\end{table}

\clearpage
\newpage
\subsection{Quantum Dynamics Simulations}

\begin{table}[h]
    \centering
    \begin{tabular}{|cc|cc|cccccccccc|}
    \hline
    \multicolumn{2}{|c|}{\multirow{2}{*}{Modes}} & \multirow{2}{*}{$N_i$} & \multirow{2}{*}{$N_j$} & \multicolumn{10}{c|}{$\text{n}_{\text{SPF}}$} \\
    &&&&GS&$S_1$&$S_2$&$S_3$&$S_4$&$T_1$&$T_2$&$T_3$&$T_4$&$T_5$\\
    \hline
    $\nu_1$    & $\nu_7$    & 41 & 41 & 1 & 18 & 12 & 12 & 12 & 2 & 2 & 2 & 2 & 2 \\
    $\nu_{11}$ & $\nu_{13}$ & 41 & 41 & 1 & 18 & 12 & 12 & 12 & 2 & 2 & 2 & 2 & 2 \\
    $\nu_{10}$ & $\nu_{12}$ & 41 & 41 & 1 & 18 & 12 & 12 & 12 & 2 & 2 & 2 & 2 & 2 \\
    $\nu_{15}$ & $\nu_{21}$ & 41 & 89 & 1 & 18 & 12 & 12 & 12 & 2 & 2 & 2 & 2 & 2 \\
    \hline
    \end{tabular}
\caption{Computational details for the MCTDH simulations of the model presented in this work. $N_i$,$N_j$ are the number of primitive harmonic oscillator discrete variable representation (DVR) basis functions used to describe each mode. $n_{\text{SPF}}$ is the number of single-particle functions used to describe the wavepacket on each state.}
\label{tab:spf}
\end{table}

\section{Trajectory Surface Hopping using ADC(2) Potentials}
ADC(2) often represents an excellent approach to perform excited state dynamics owing to its favourable balance between computational efficiency and accuracy as well as its increased stability compared to methods such as CC2 \cite{plasser2014surface}. However, as discussed in ref. \cite{marsili2021caveat} and shown above in Figure S2, it shows artificial crossing between the ground and first excited state along the C=O stretch. To understand the influence this has on the excited state dynamics and the ultrafast electron diffraction, we have performed excited state dynamics dynamics using ADC(2) potentials, and the results are now discussed. 

\begin{figure}[h]
    \centering
    \includegraphics[width=0.6\linewidth]{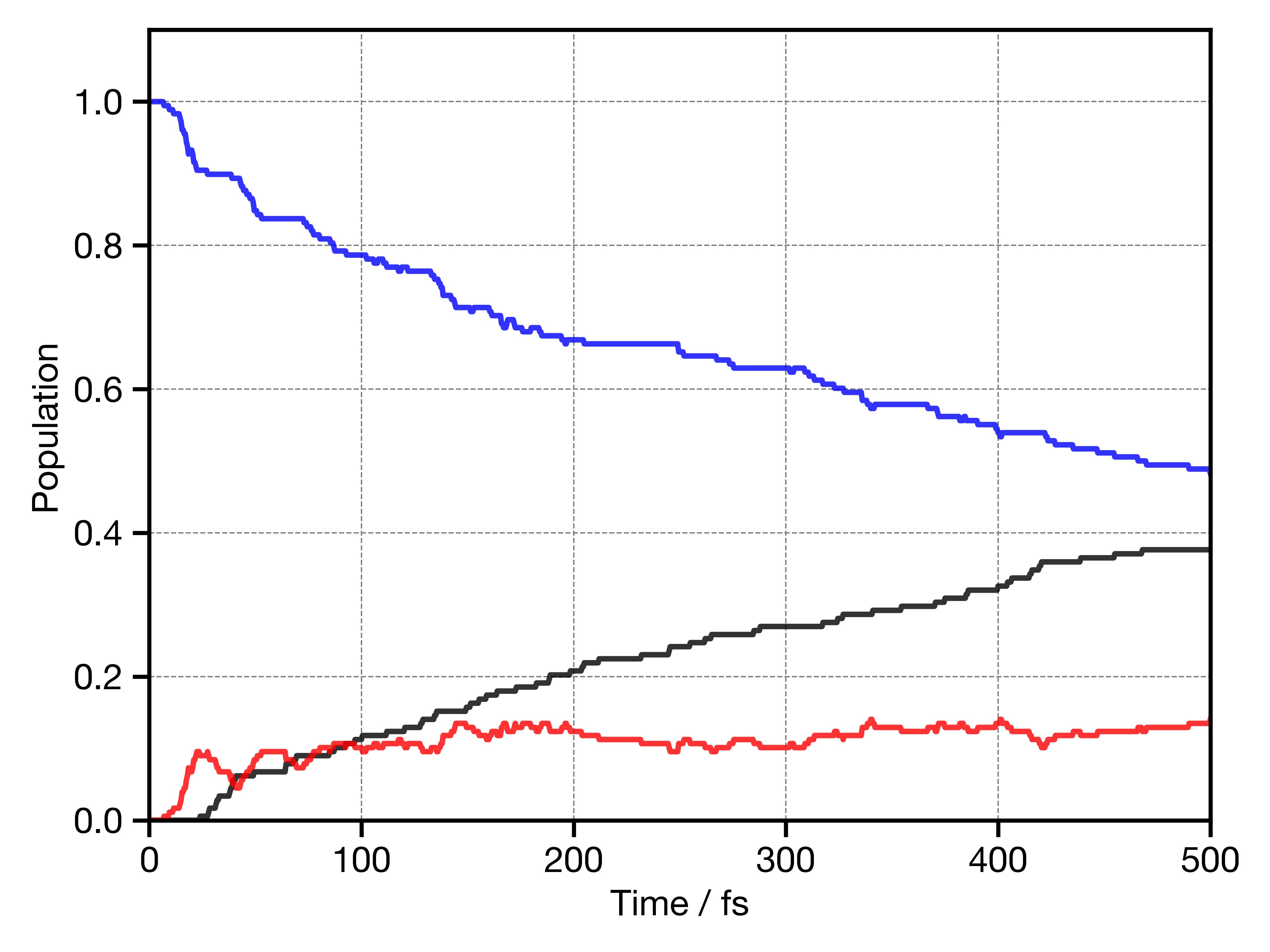}
    \caption{Excited state population kinetics extracted from 69 \textit{on-the-fly} trajectory surface hoping trajectories after vertical excitation into the S$_2$ (blue) following the decay into the S$_1$ (red) and ground (black) state. All trajectories included were propagation for 500fs.}
    \label{fig:TSHpopkinetics}
\end{figure}

Figure \ref{fig:TSHpopkinetics} shows the relative populations of the S$_2$, S$_1$ and ground state for the first 500 fs after excitation into the S$_2$(n$\rightarrow$3s) state obtained from 69 trajectories, ran for 500 fs. This shows a comparable, but slightly slower decay of the S$_2$ state than observed for LR-TDDFT(PBE0) and the quantum dynamics shown in the main text. Indeed, this decay kinetics is actually in very close agreement with the photoelectron spectroscopic study performed in ref.  \cite{kuhlman2012coherent}. In agreement with the LR-TDDFT(PBE0), once in the S$_1$(n$\rightarrow\pi^{\ast}$) (red) state, the trajectories relax quickly into the electronic ground state (black) with a constant amount of $\sim$15\% remaining in the S$_1$ state.

\begin{figure}[h]
    \centering
    \includegraphics[width=0.9\linewidth]{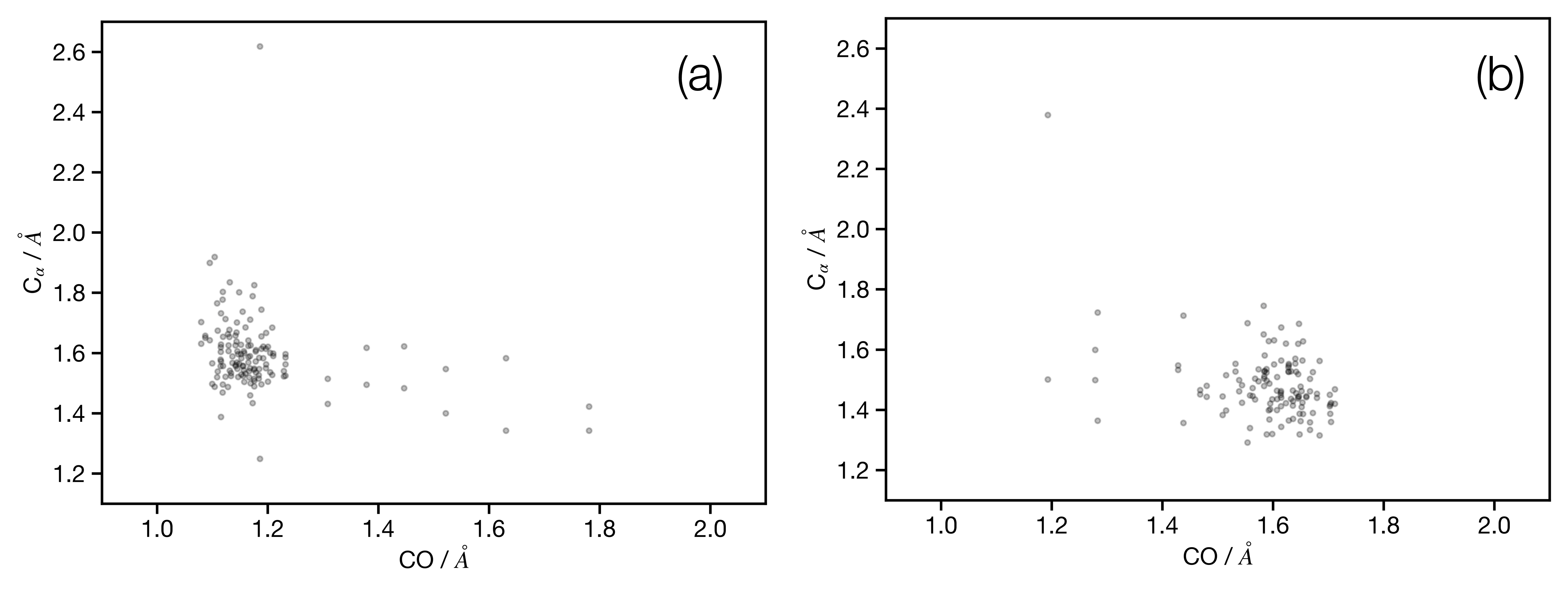}
    \caption{Average C$_{\alpha}$ and CO bond lengths for the structure of cyclobutanone where the trajectory hops from S$_2$-S$_1$ (a) and S$_1$-S$_0$ (b).}
    \label{fig:hoppingpoint}
\end{figure}

Figure \ref{fig:hoppingpoint} shows an analysis of the hopping point of all the trajectories. This shows, consistent with the LR-TDDFT(PBE0) and model Hamiltonian dynamics presented in the main text that conversion from the S$_2\rightarrow$S$_1$ occurs close to the Franck-Condon geometry. This fact that the dynamics is slow, despite the small structural displacements required to reach the crossing geometries reflects the weak nature of the coupling, which to first order is symmetry forbidden. In contrast, the hopping geometries of the conversion from the  S$_1\rightarrow$S$_0$ state demonstrates a larger displacement, primarily when the C=O distance is 1.6 \mbox{\AA}.  As discussed above, is the point for the artificial crossing between the S$_1$ and S$_0$ state, which arises along the C=O stretch for ADC(2), but not LR-TDDFT(PBE0) or NEVPT2. The importance of this false crossing is clearly likely to bias the excited state dynamics simulations meaning that crossing to the ground state will more likely occur close to the Franck-Condon geometry with a elongated C=O bond rather than the ring open conformers identified in during previous work \cite{liu2016new}. Indeed, this bias can be seen in the formation of the photoproducts, with only 7\% of the trajectories exhibiting dissociative behaviour.

\begin{figure}[h]
    \centering
     \includegraphics[width=1.0\linewidth]{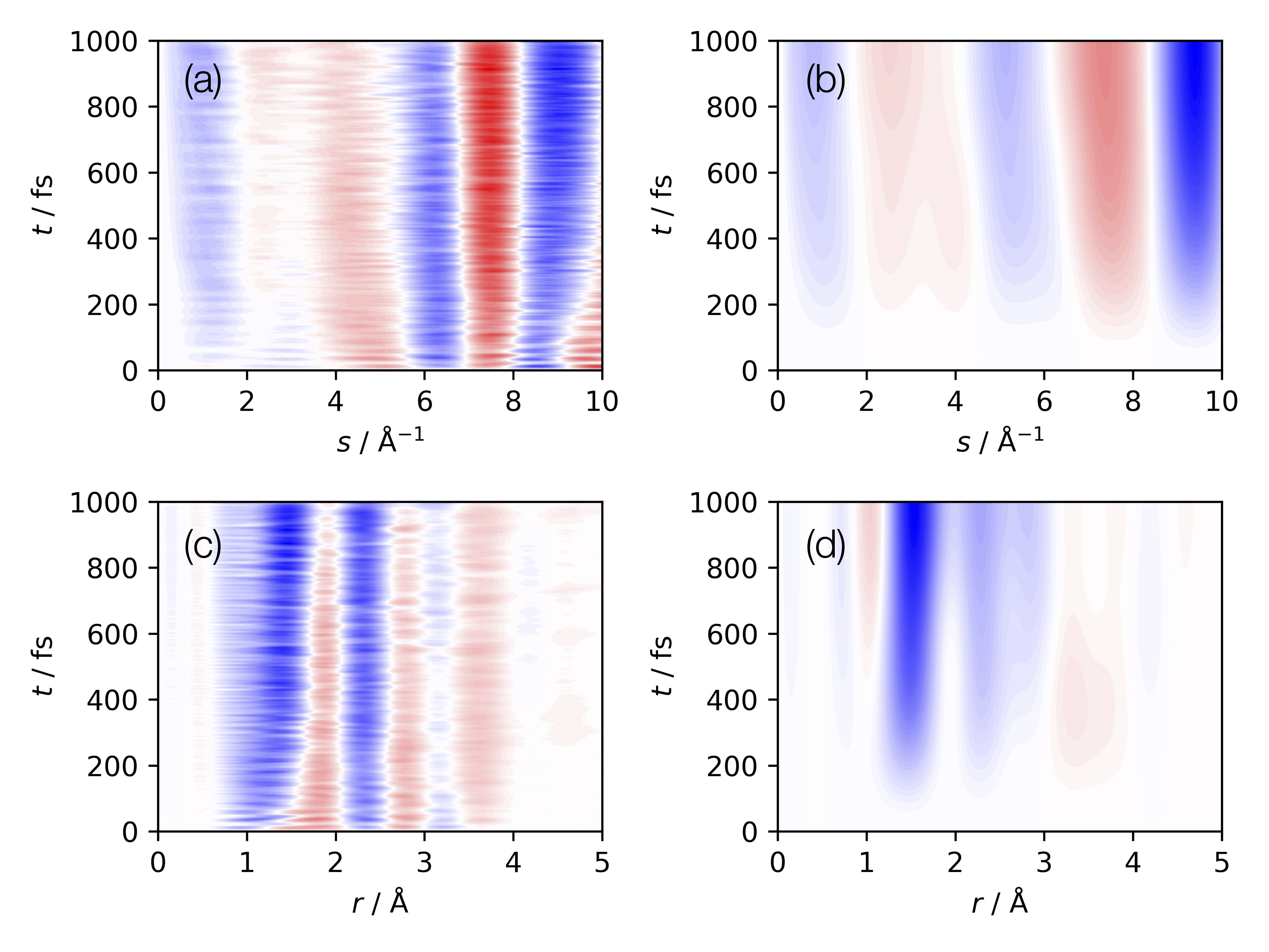}\\
    \caption{Transient scattering ($\Delta I/I$) without (a) and with (b) 150 fs temporal broadening. Transient PDF without (c) and with (d) 150 fs temporal broadening. The ground state curves used to generate the transient were obtained from from the initial conditions, \textit{i.e.} at t=0. All these plots have been calculated using the 69 500 fs trajectories.}
    \label{fig:trelectrondiffraction}
\end{figure}

Figure \ref{fig:trelectrondiffraction} shows the time resolved electron diffraction simulations arising after photoexcitation of cyclobutanone into the S$_2$ state. Figure \ref{fig:trelectrondiffraction}a shows the raw scattering signal, while Figure {fig:trelectrondiffraction}b has been broadened with a Gaussian with width of 150 fs, consistent with the temperoal resolution of the experiment. Figures \ref{fig:trelectrondiffraction}c and d show the sine transformation in the time-resolved PDF without and with temporal broadening, respectively.  The time-resolved scattering curves exhibit broadly similar behaviour to the LR-TDDFT(PBE0) shown in the main text with strong  2 strong negative ($\sim$1 and 9 \mbox{\AA}$^{-1}$) and 2 positive ($\sim$2.7 and 7.5 \mbox{\AA}$^{-1}$) features. Importantly, each of the transient features exhibits a slight movement to small distances with time, suggesting an elongation of the average bond distances consistent with those trajectories that dissociate, but also the C=O bond changes discussed above. 

\begin{figure}[h]
    \centering
     \includegraphics[width=0.6\linewidth]{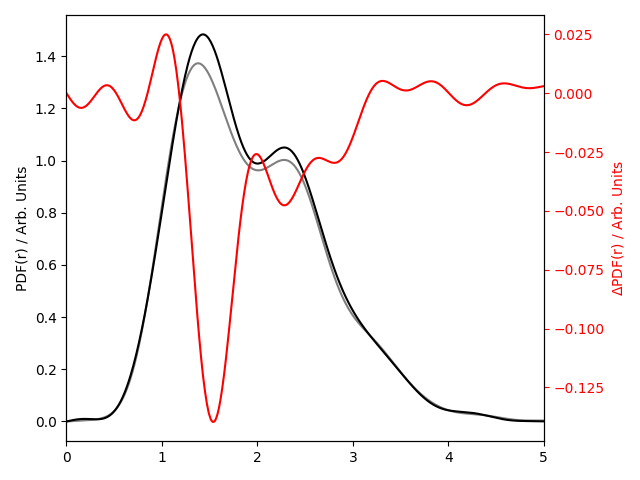}\\
    \caption{Initial (t=0, black) and final (t=2000 fs, grey) PDF and their difference (red) calculated using the 289 2000 fs surface hopping  trajectories simulated using potentials are ADC(2) level of theory.}
    \label{fig:pdf}
\end{figure}

The time-resolved PDFs (Figures \ref{fig:trelectrondiffraction}c and d and Figure \ref{fig:pdf}) also shows 2 negative peaks at 1.5 and 2.5 \mbox{\AA}, consistent with the LR-TDDFT(PBE0) dynamics. However, the relative ratio between these peaks is rather different, with the negative transition at 1.5 \mbox{\AA} being significantly stronger than the one at 2.5 \mbox{\AA}. This arises because although the fragmentation leads to the loss of both features, in the present dynamics,  the much stronger changes in the C=O bond length moves intensity from the first. to the second peak, making the change in the first peak larger. 

\clearpage
\newpage

\section{Key Geometries}

\begin{table}[ht]
  \centering
  \label{tab:coordinates}
  \begin{tabular}{c|ccc}
    \hline
    Atom & $x$ (\AA) & $y$ (\AA) & $z$ (\AA) \\
    \hline
C &    -0.22  &   0.64   &  0.00  \\
C &    -0.11  &  -0.41   &  -1.10 \\
C &     0.24   &  -1.45  &   0.00  \\
C &    -0.11  &  -0.41   &   1.10  \\
O &    -0.28   &   1.84   &   0.00  \\
H &     0.63   &  -0.21  &  -1.88 \\
H &    -1.08   &  -0.59  &  -1.58 \\
H &    -0.35   &  -2.36  &   0.00  \\
H &     1.31  &  -1.71   &  0.00  \\
H &     0.63   &  -0.21  &   1.88  \\
H &    -1.08   &  -0.59  &   1.58 \\
    \hline
  \end{tabular}
\caption{Ground state geometry optimised using DFT(PBE0)/aug-cc-pvdz.}
\end{table}

\begin{table}[ht]
  \centering
  \label{tab:coordinates}
  \begin{tabular}{c|ccc}
    \hline
    Atom & $x$ (\AA) & $y$ (\AA) & $z$ (\AA) \\
    \hline
C  &    0.04  &    0.70  &  0.00  \\
C  &   -0.06  &   -0.42  &  -1.08  \\
C  &    0.25 &   -1.47  &   0.00  \\ 
C  &  -0.06   &  -0.42   &  1.08   \\
O  &   -0.59 &    1.80  &   0.00  \\
H  &    0.64  &   -0.28   &  -1.91  \\
H  &   -1.10 &   -0.48  &  -1.46  \\
H  &   -0.39 &   -2.35   &   0.00  \\
H  &    1.30 &   -1.77   &  0.00  \\
H  &    0.64  &   -0.28   &   1.91  \\
H  &   -1.10 &   -0.48  &   1.46  \\
    \hline
  \end{tabular}
\caption{Excited S$_1$ excited state geometry optimised using LR-TDDFT(PBE0)/aug-cc-pvdz.}
\end{table}

\begin{table}[ht]
  \centering
  \label{tab:coordinates}
  \begin{tabular}{c|ccc}
    \hline
    Atom & $x$ (\AA) & $y$ (\AA) & $z$ (\AA) \\
    \hline
  C &  1.61   &  0.00    &   0.00  \\
  C &  0.52    &   1.06    &   0.00 \\
  C &  0.52    &  -1.06    &   0.00 \\
  C &  -0.64   &   0.00   &   0.00 \\
  O &  -1.80   &    0.00   &   -0.01 \\
  H &  2.22   &  0.00   &   0.91 \\
  H &  2.22    &  0.00    &  -0.90 \\
  H &  0.39    &   1.62    &   0.94 \\
  H &  0.40   &   1.63   &  -0.94 \\
  H &  0.40   &  -1.63   &  -0.94 \\
  H &  0.39    &  -1.63   &   0.94 \\
    \hline
  \end{tabular}
\caption{Excited S$_2$ excited state geometry optimised using LR-TDDFT(PBE0)/aug-cc-pvdz.}
\end{table}

\begin{table}[ht]
  \centering
  \label{tab:coordinates}
  \begin{tabular}{c|ccc}
    \hline
    Atom & $x$ (\AA) & $y$ (\AA) & $z$ (\AA) \\
    \hline
C   &  -0.31   &   0.64   &  -0.09  \\
C   &   0.16  &  -0.43   &  -1.04 1 \\
C   &   0.27   &  -1.53  &  -0.01 \\
C   &  -0.12   &  -0.48   &   1.02  \\
O   &  -0.51  &   1.80  &  -0.06 \\
H   &   0.89  &  -0.12   &  -1.80  \\
H   &  -1.18   &  -0.41   &  -1.16 \\
H   &  -0.52  &  -2.30  &  -0.13 \\
H   &   1.26  &  -1.96   &   0.13 \\
H   &   0.73  &  -0.07   &   1.59  \\
H   &  -1.09  &  -0.58   &   1.54  \\
    \hline
  \end{tabular}
\caption{S$_2$/S$_1$ conical intersection geometry optimised using LR-TDDFT(PBE0)/aug-cc-pvdz.}
\end{table}

\begin{table}[ht]
  \centering
  \label{tab:coordinates}
  \begin{tabular}{c|ccc}
    \hline
    Atom & $x$ (\AA) & $y$ (\AA) & $z$ (\AA) \\
    \hline
C  &   -0.13  &   0.87  &   0.26 \\
C  &   -0.23  &  -0.82  &  -1.38 \\
C  &    0.26  &  -1.47  &  -0.13 \\
C  &   -0.07  &  -0.44  &   0.95 \\
O  &   -0.61  &   1.92  &   0.12 \\
H  &    0.29  &   0.11  &  -1.67 \\
H  &   -0.62  &  -1.41  &  -2.21 \\
H  &   -0.28  &  -2.40  &   0.08 \\
H  &    1.34  &  -1.72  &  -0.10 \\
H  &    0.65  &  -0.34  &   1.79 \\
H  &   -1.07  &  -0.65  &   1.37 \\
    \hline
  \end{tabular}
\caption{S$_1$/S$_0$ C$_{\alpha}$ conical intersection geometry optimised using LR-TDDFT(PBE0)/aug-cc-pvdz.}
\end{table}

\begin{table}[ht]
  \centering
  \label{tab:coordinates}
  \begin{tabular}{c|ccc}
    \hline
    Atom & $x$ (\AA) & $y$ (\AA) & $z$ (\AA) \\
    \hline
C  &    0.07   &   0.55  &   0.04 \\
C  &    0.66   &  -0.42  &  -1.17 \\
C  &    2.04  &  -0.23   &  -1.48 \\ 
C  &   -0.17  &  -0.11   &   1.25 \\
O  &   -0.10   &   1.72  &  -0.25  \\
H  &    0.00   &  -0.12   &  -1.99  \\
H  &    0.43   &  -1.44  &  -0.81  \\
H  &    2.82  &  -0.78   &  -0.94  \\
H  &    2.36   &   0.50  &  -2.23 \\
H  &    0.01   &  -1.18  &   1.36  \\
H  &   -0.40   &   0.49  &   2.12  \\
    \hline
  \end{tabular}
\caption{S$_1$/S$_0$ C$_{\beta}$ conical intersection geometry optimised using LR-TDDFT(PBE0)/aug-cc-pvdz.}
\end{table}

\begin{table}[ht]
  \centering
  \label{tab:coordinates}
  \begin{tabular}{c|ccc}
    \hline
    Atom & $x$ (\AA) & $y$ (\AA) & $z$ (\AA) \\
    \hline
C  &   -0.46   &   0.72  &  -0.26  \\
C  &    0.78  &  -1.60  &  -1.92  \\
C  &    0.40  &  -2.69  &  -1.23 \\
C  &    0.43  &  -0.31  &   0.36 \\
O  &   -1.66   &   0.49  &  -0.26 \\
H  &    1.83  &  -1.33  &  -2.02 \\
H  &    0.05   &  -1.03  &  -2.50 \\
H  &   -0.65  &  -2.97   &  -1.14 \\
H  &    1.13  &  -3.27  &  -0.64  \\
H  &    1.44  &   0.02  &   0.61 \\
H  &    0.13  &  -1.32  &   0.62 \\
    \hline
  \end{tabular}
\caption{S$_1$/S$_0$ C$_{\alpha}$, C$_{\beta}$ conical intersection geometry optimised using LR-TDDFT(PBE0)/aug-cc-pvdz.}
\end{table}